# Picosecond volume expansion drives a later-time insulator-metal transition in a nano-textured Mott Insulator

Anita Verma[1], Denis Golež[2,3], Oleg Yu. Gorobtsov[1], Kelson Kaj[4], Ryan Russell[5], Jeffrey Z. Kaaret[6], Erik Lamb[4], Guru Khalsa[1], Hari P Nair[1], Yifei Sun[1], Ryan Bouck[1], Nathaniel Schreiber[1], Jacob P. Ruf[7], Varun Ramaprasad[4], Yuya Kubota[8,9], Tadashi Togashi[8,9], Vladimir A. Stoica[10], Hari Padmanabhan[10], John W. Freeland[11], Nicole A. Benedek[1], Oleg Shpyrko[4], John W. Harter[5], Richard D. Averitt[4], Darrell G. Schlom[1,12,13], Kyle M. Shen[7,12], Andrew J. Millis[14,15], and Andrej Singer[1*]

[1]*Department of Materials Science and Engineering, Cornell University, Ithaca, New York 14853, USA*
[2]*Jozef Stefan Institute, Jamova 39, SI-1000 Ljubljana, Slovenia*
[3]*Faculty of Mathematics and Physics, University of Ljubljana, Jadranska 19, 1000 Ljubljana, Slovenia*
[4]*Department of Physics, University of California San Diego, La Jolla, California 92093, USA*
[5] *Materials Department, University of California, Santa Barbara, California 93106, USA*
[6]*School of Applied and Engineering Physics, Cornell University, Ithaca, New York 14853, USA*
[7]*Physics Department, Cornell University, Ithaca, New York 14853, USA*
[8] *RIKEN SPring-8 Center, 1-1-1 Kouto, Sayo-cho, Sayo-gun, Hyogo 679-5148, Japan*
[9] *Japan Synchrotron Radiation Research Institute, 1-1-1 Kouto, Sayo-cho, Sayo-gun, Hyogo 679-5198, Japan*
[10] *Department of Materials Science and Engineering, Pennsylvania State University, University Park, PA, USA*
[11] *X-ray Science Division, Argonne National Laboratory, Argonne, Illinois 60439, USA*
[12] *Kavli Institute at Cornell for Nanoscale Science, Cornell University, Ithaca, NY 14853, USA*
[13] *Leibniz-Institut für Kristallzüchtung, Max-Born-Straße 2, 12489 Berlin, Germany*
[14] *Center for Computational Quantum Physics, Flatiron Institute, 162 5th Avenue, New York, NY 10010, USA*
[15] *Department of Physics, Columbia University, 538 West 120th Street, New York, NY 10027, USA*
*asinger@cornell.edu

**Technology moves towards ever faster switching between different electronic and magnetic states of matter. Manipulating properties at terahertz rates requires accessing the intrinsic timescales of electrons (femtoseconds) and associated phonons (10s of femtoseconds to few picoseconds), which is possible with short-pulse photoexcitation. Yet, in many Mott insulators, the electronic transition is accompanied by the nucleation and growth of percolating domains of the changed lattice structure, leading to empirical time scales dominated by slow coarsening dynamics. Here, we use time-resolved X-ray diffraction and reflectivity measurements to investigate the photoinduced insulator-to-metal transition in an epitaxially strained thin film Mott insulator $Ca_2RuO_4$. The dynamical transition occurs without observable domain formation and coarsening effects, allowing the study of the intrinsic electronic and lattice dynamics. Above a fluence threshold, the initial electronic excitation drives a fast lattice rearrangement, followed by a slower electronic evolution into a metastable non-equilibrium state. Microscopic calculations based on time-dependent dynamical mean-field theory and semiclassical lattice dynamics within a recently published equilibrium energy landscape picture explain the threshold-behavior and elucidate the delayed onset of the electronic phase transition in terms of kinematic constraints on recombination. Analysis of satellite scattering peaks indicates the persistence of a strain-induced nano-texture in the photoexcited film. This work highlights the importance of combined electronic and structural studies to unravel the physics of dynamic transitions and elucidates the role of strain in tuning the timescales of photoinduced processes.**

**Introduction**

   The multiband Mott insulator $Ca_2RuO_4$ is a layered perovskite oxide with significant electronic correlations[1]. It exhibits a high-temperature metallic phase and a low-temperature insulating phase[2]. The temperature-driven insulator-metal transition in $Ca_2RuO_4$ is accompanied by an isosymmetric structural transition: the symmetry is *Pbca* in both phases, yet the 'c-axis' changes from short (insulating phase, *S-Pbca*) to long (metallic phase, *L-Pbca*) with a coincident decrease in the tilt of the $RuO_6$ octahedra[3] (see Fig. 1a). The relevant electronic states are derived from the Ru 4d shell, which contains 4 electrons per Ru. In the insulating phase, the lower energy $d_{xy}$ orbital is doubly filled while the higher energy orbitals, $d_{yz}$, and $d_{xz}$ are half-filled[4-7]. In the metallic phase, the $d_{xy}$, $d_{yz}$, and $d_{xz}$ orbitals are each approximately 4/3 filled (see Fig. 1a)[8]. These octahedral modes are strongly coupled to global strain fields[9]. The equilibrium properties of the material strongly correlate with its local crystal structure, especially with the rotation, tilting, and flattening of the $RuO_6$ octahedra. The insulating phase and the metal-insulator transition are highly susceptible to external perturbations such as temperature, pressure, electric field, and chemical substitution, enabling the control of magnetic, structural, transport, and physical properties from a technological application perspective[10-13].

Ultrafast photoexcitation opens a new paradigm for controlling materials properties. In addition to elucidating fundamental interactions between various degrees of freedom out-of-equilibrium[14], photoexcitation offers opportunities for future electronics in the terahertz range and beyond, providing the possibility of switching between two states on an intrinsic timescale not associated with moving carriers into or out of a device[15]. Many Mott insulator materials exhibit a crystal symmetry change across the insulator-metal transition. In these systems, nucleation and growth of structural domains following the Avrami model[16] can significantly slow down the insulator-to-metal transition[17-19]. Here, we use short optical laser pulses to induce an ultrafast isostructural insulator-metal transition in biaxially strained $Ca_2RuO_4$ thin films, where the isostructural nature of the transition suggests that domain effects may be avoided. We monitor the physics by a combination of time-resolved femtosecond X-Ray diffraction and broad-band reflectivity, enabling a comparative study of the lattice and electronic dynamics. We demonstrate a photoinduced 4 ps lattice expansion, followed by a slower increase in the low-frequency reflectivity. We use time-dependent dynamical mean-field theory to link the experimental data to Mott physics in the presence of coupling to the lattice. The experiment and theory both indicate a photoinduced transition into a non-thermal state.

**Photoinduced lattice expansion and insulator-to-metal transition**

We apply short optical laser pulses ($E_{pump} = 1.55\ eV$ duration 40 fs) to 25 nm thick films of $Ca_2RuO_4$ epitaxially grown on $LaAlO_3$ [001] substrates (see Methods). The films were held at 130 K, well below the equilibrium metal-insulator transition temperature of 200K (lower than the bulk metal-insulator transition temperature[2] of 357 K due to epitaxial strain from substrate clamping). The pump excites carriers across the 1.2 eV band gap of insulating $Ca_2RuO_4$. The penetration depth of the pump field is greater than the thickness of the sample, so we may assume the entire sample volume is uniformly excited. We interrogate the photoinduced dynamics with femtosecond X-Ray diffraction to probe the lattice and time-resolved high and low-frequency reflectivity to probe the electronic state (see Methods).

As a reporter of the lattice dynamics, we choose the 008 Bragg peak, related to the 'c-axis' lattice parameter. This peak is visible near momentum transfers of $q_z = (4\pi/\lambda)\sin(\theta)$, where $\lambda = 1.26$ Å is the X-ray wavelength and $\theta = 25$ ° is the incident angle (see inset in Fig. 1b). The main panel of Fig. 1b shows snapshots of the scattering intensity for wavevectors near this peak for a large amplitude photoexcitation (50 mJcm$^{-2}$ incident fluence). By 4 ps after photoexcitation, the intensity maximum has shifted to a lower momentum transfer corresponding to a photoinduced expansion of the 'c-axis', consistent with a transition from *S-Pbca* (insulating) to *L-Pbca* (metallic) phase (see Fig. 1a,b): the expansion persists for long times after the excitation (more than 49 ps). The black squares in Fig. 1c show the intensity at the wave vector of the peak in the long-time state; the lattice dynamics exhibit a 1.2±0.6 ps rise time (see Fig. S2), persist up to the maximum measured 100 ps, and display small oscillations about the long-time value. The few-ps lattice expansion timescale observed here is consistent with a strain wave propagating through the film as a coherent acoustic phonon[20] (see Supplementary Material and Fig. S2).

Fig. 1c presents two reporters of the electronic state. The red circles show the time-resolved reflectivity at a probe frequency of 1.55 eV, relevant to the rearrangement of the high-energy, above-band-gap excitations. At this energy, we observe a rapid photoinduced change in the electronic properties (time delays, $\tau$ < 1 ps, faster than the lattice response) followed by a peak in reflectivity coincident with the time of the lattice expansion and a slow decrease over the 100 ps measurement timescale, indicating that the pump has produced a sudden and long-lived change in the electronic state. The purple triangles show the time-resolved reflectivity in the THz frequency range (1 meV – 10 meV), which indicates a possible metallic response of the photoinduced carriers. This signal has a significantly slower rise time than both the high-frequency reflectivity and the lattice expansion, and no decay is evident in the 100 ps measurement window.

**Photoinduced atomic rearrangements within the unit cell**

In addition to the lattice expansion discussed above, the time-resolved X-ray diffraction provides snapshots of the local structure through the unit cell structure factor, $F_{hkl}$, accessible via the integrated peak intensity[21] $I_{hkl} = |F_{hkl}|^2$, which is the shaded area under the curves in Fig. 1b. In the 008 Bragg condition for $Ca_2RuO_4$, the Ru atoms scatter in-phase, but the neighboring calcium layers and the apical oxygen layers scatter almost exactly out-of-phase, generating partial destructive interference (see Fig. 2a). As a result, the structure factor magnitude $|F_{008}|$ is highly susceptible to slight changes in the Wyckoff positions of the atoms. In equilibrium, $I_{008}$ is more than twice higher in the high-temperature metallic phase than in the low-temperature insulating phase, in large part because thermal disorder leads to deviations from the destructive interference condition (see Supplementary Figs. S3-S6 for density functional theory showing qualitative agreement with quasi-static experimental data).

Fig. 2b presents our data of the change in 'c-axis' lattice parameter, $\Delta c$, against change in the integrated peak intensity, $\Delta I_{008}$, for different fluences at different time-delays (see also Figs. S7-S8); we also show results obtained in equilibrium at different temperatures. After photoexcitation at lower fluences, the lattice spacing remains practically unchanged, $\Delta c / c_{GS} <$ 0.05 %, while the magnitude of the integrated peak intensity increases, albeit to a value less than the high-temperature equilibrium value. Photoexcitation at a high fluence drives the lattice expansion to a value comparable to the value measured in equilibrium at 573 K, above the transition temperature of 200 K, while the intensity $\Delta I_{008}$ still remains far below the metallic-phase equilibrium value. This comparison of $\Delta I_{008}$ and $\Delta c$ strongly suggests the existence of a fluence threshold[22] (see Fig. S9), below which the photoexcitation excites the system around the local, insulating state, but is insufficient to drive a dynamical transition to the long-bond state reminiscent of L-Pbca. The comparison also indicates that the high fluence state, which persists for at least 50 ps (see Fig. 2b), is non-thermal, exhibiting lattice parameters and transmission properties similar to the metallic state but with a local fluctuation amplitude (local lattice temperature) that is more similar to the low-temperature phase.

**Theory**

Ca$_2$RuO$_4$ involves a correlated electron fluid coupled to lattice distortions. The dominant electron-electron interactions are the on-site Slater-Kanamori (U-J) multiplet interactions. The lattice distortions involve changes to the lattice constant, shape, and orientation of the Ru-O$_6$ octahedra. Because the lattice response is approximately harmonic[23], we may express the electron-phonon coupling in terms of the dominant unit cell normal mode $Q_3 = \left(\frac{1}{\sqrt{6}}\right) \cdot (2c - a_0 - b_0)$ [23,24], where a, b, and c are the lattice constants of the approximately tetragonal unit cell, and express the lattice energetics in terms of a distortion energy $\frac{1}{2}KQ_M^2$, where K is an effective elastic constant obtained by integrating out all of the other modes except $Q_3$ within a harmonic approximation in the presence of the epitaxial constraint imposed by the substrate and $Q_M$ is the value of $Q_3$ in the equilibrium low-temperature insulating state. In the thin film geometry of interest here, the in-plane lattice constants $(a, b)$ are, to good accuracy[23], fixed by the epitaxial constraints to the substrate values $(a_0, b_0)$, while the out of plane lattice parameter, $c$, changes from a longer to a shorter value as the transition line is crossed from metal to insulator (see Methods).

We employ time-dependent Dynamical Mean-Field Theory[25,26] for the electronic states and treat the lattice dynamics on a semiclassical level using model parameters from previous work[8] (see Methods for details). The fundamental calculational bottleneck is the limited propagation time available for treating the electronic problem with state-of-the-art methods[27]. The initial photoexcitation is accessible, but capturing the dynamics even over one optical phonon oscillation period is challenging. Due to the limited propagation time, we artificially increase the frequency, $\omega_0$, of the lattice mode, $Q_3$, from the physical values of 0.075 eV to 0.25 eV so that the calculation can resolve several phonon cycles[28-30] (see Supplementary Materials and Fig. S10 for scaling to the low-frequency regime). We perform calculations out to the longest computationally accessible times and interpret the results within the theoretical structure established for the equilibrium problem.

We describe the results in terms of the coupled order parameter theory introduced in ref. [8]. We use the difference in occupancy of the different Ru d-orbitals to define the electronic order parameter as $\phi = n_{xy} - (n_{xz} + n_{yz})/2$ with $\phi=1$ ($\phi=0$) for insulating (metallic) phase[1,4-7,23] and use the unit cell normal mode, $Q_3$, for the lattice order parameter. Results are shown in Fig 3. The initial photoexcitation promotes electrons from the $d_{xy}$ orbital to the upper Hubbard band of $d_{yz}/d_{xz}$ orbitals. We parametrize the photoexcitation strength by the number of carriers $dn_1$ excited out of the xy-band and into the other two orbitals. The excited carriers are found to relax very rapidly (few fs) to a pseudo-equilibrium state in which the excited electrons and holes thermalize within their many-body bands. The number of excited particles is approximately conserved because recombination to true equilibrium is found to be strongly suppressed by kinetic barriers and for fixed lattice positions typically occurs only on picosecond timescales[31-34], far outside of our computational window.

The initial excitation rapidly reduces the electronic order parameter (see Fig. 3b). The direct coupling between the electronic order parameter $\phi$ and the unit cell distortion $Q_3$ creates, in effect, an impulsive force on the lattice mode, which induces the oscillations and a shift of the mean position shown in Fig. 3a. At intermediate photoexcitation ($dn_1 =\leq 4\%$), the mean position is only slightly different from the initial equilibrium position and remains steady in time while the oscillation amplitude slowly decays. At stronger excitation, the Figs. 3a,b reveal a long-time drift in both the mean lattice distortion, $Q_3$, and the electronic order parameter, $\phi$, towards the metallic solution: at its largest amplitude, the oscillating lattice distortion $Q_3$ approaches the value $Q_3=0$ corresponding to the metallic state. Over the time range accessible to our calculations, we may parametrize the time dependence of the order parameter by fitting the oscillating curves to a straight line with slope $\dot\phi$. Fig. 3c presents the slope as a function of photodoping excitation strength $dn_1$. A threshold-type behavior is observed, with a zero drift below the critical excitation and a nonzero drift for stronger excitations, consistent with the trends observed in $Q_3$ (see Fig. 3a). The calculated fluence threshold is reduced with the decreasing lattice energy $KQ_m^2/2$, see Fig. 3c (see Fig. S11). In Ref.[8], it was shown that the lattice energy $KQ_m^2/2$ is reduced with the increasing epitaxial strain (see Supplementary Material for analysis).

To interpret the results and extend them to times beyond the scales accessible to the time-dependent dynamical mean-field calculations, we consider the free energy landscape shown in Fig. 3d[8] that is a functional of an electronic order parameter $\phi$ and a lattice coordinate $Q_3$. The strong coupling between the electronic and lattice orders in this material is reflected in the structure of the landscape, which features two local minima, at $\phi = Q_3 = 0$ (metal) and at $\phi = 1; Q_3 = Q_m$ (insulator), while configurations with only electronic ordering $\phi = 1; Q_3 = 0$ or only lattice distortions $\phi = 0; Q_3 = Q_m$ correspond to local maxima in the equilibrium landscape.

We present the time-dependent DMFT results as a system trajectory in the free energy landscape of Fig. 3d. The system starts in the low-temperature minimum ($\phi \approx 1; Q_3 \approx Q_m$). The initial photoexcitation changes the electronic order parameter on a very rapid timescale at which the lattice cannot respond, in effect moving the system horizontally to the left in the energy landscape. If the photoinduced change in the electronic order parameter, $\phi$, is small, the system remains within the basin of attraction of the low-temperature phase (red trace), and the result is oscillations gradually decaying back to equilibrium but no long-term drift. If the photoinduced change is large enough (cyan and green traces), the photoexcitation moves the system beyond the basin of attraction of the insulating phase, and the dynamics is ultimately a relaxation to the minimum corresponding to the high-temperature state. Our calculation does not permit access to the multi-ps times needed to resolve the long-time evolution. Nevertheless, we can observe that the kinetic barrier to electronic relaxation found in our calculation means that the first stages after the initial excitation correspond to lattice oscillations driven by a weakly time-dependent electronic disproportionation, with further evolution of the electronic state only occurring as the lattice distortion evolves to smaller values. Further, the relaxation of the lattice to its final state is constrained by elastic issues, which require that strain is removed via the propagation of acoustic waves into the substrate, requiring a time of the order of the film thickness divided by the sound velocity (see supplementary material). Despite the uncertainties, the theoretical modeling reveals that the origin of the principal features of the experimental findings is a partially unexpected motion of the system in an energy landscape, with an excitation above a critical strength required to move the system out of the basin of attraction of the equilibrium state and the subsequent dynamics being controlled by lattice timescales because of the kinetic barrier to electron relaxation.

**Collective lattice vibrations and non-equilibrium nano-texture**

The diffraction experiments contain additional information that provides a more nuanced view of the dynamical phase transition. By scanning the angle of the incident X-ray and recording sections of the Ewald sphere with an area detector[35] we collected 3D reciprocal space data (see Fig. 4a for a slice in the $q_x$-$q_z$ plane, and Figs. S12-S14). The bright center spot is the 008 Bragg peak of $Ca_2RuO_4$, elongated due to the finite film thickness along the momentum transfer direction $q_z$ normal to the film surface[21]. In equilibrium, multiple orders of satellite peaks are visible horizontally off-center in the low-temperature ground state, which persist down to the lowest measured temperature of 7 K[36]. The satellites are confined to the planes along $q_x$-$q_z$ and $q_y$-$q_z$ and reveal the presence of a nano-texture with well-defined periodicity and crystallographic orientation, akin to ferroelastic domains in ferroelectric materials[37,38]. Real-space imaging in similarly synthesized $Ca_2RuO_4$/$LaAlO_3$ films shows alternating diagonally oriented, approximately 9 nm wide layers with larger and smaller out-of-plane lattice constants[36]. Despite the nano-texture, the time-resolved THz data and resistivity measurements (see Supplementary

Fig. S15) suggest that the low-temperature state is insulating (with about 20% metallic fraction). Therefore, our discussion of Figs. 1-3 is to zeroth order unaffected by the nano-texture.

Photoexcitation modifies the intensity of the satellites (see Figs. 4 (b-c), Figs. S16-S18), indicating photoinduced dynamics of the nano-texture. The satellite intensities oscillate, revealing that photoexcitation launches long-wavelength coherent vibrations in the system with the frequency depending on the distance from the Bragg peak (see Fig. 4d and Figs. S17-S19). The second-order satellites at $|q|=0.03$ Å$^{-1}$ oscillate for 10 ps, and the first-order satellites at $|q|=0.015$ Å$^{-1}$ oscillate for 50 ps (see Fig. 4d). A Fourier analysis of the time-resolved data yields the dispersion relation of the photoinduced lattice vibrations with an in-plane wavevector along $q$ (see Fig. 4e and Fig. S20). The analysis resembles calculating the phonon dispersion in Fourier transform inelastic X-ray scattering[39,40]. In contrast to squeezed phonons[39,40] with a monotonic variation of Fourier components as a function of wavevector, here, the lattice vibrations are centered at $q = 0.015$ Å$^1$, the wavevector of the satellite peaks in equilibrium (see Fig. 4f), similar to lattice vibrations driven in ordered systems such as charge density waves[41,42]. Fourier components below 0.01 Å$^{-1}$ are strongly suppressed, suggesting no vibrations at a period larger than 60 nm.

The shape and intensity of the satellite peaks capture the spatial arrangement of the nano-texture. At 100 ps after photoexcitation at the highest fluence, the X-ray data still show the first-order satellite peaks (absent in the high-temperature equilibrium phase), revealing persistent nano-texture after photoexcitation. We have not measured time delays larger than 100 ps; nevertheless, we see no appreciable decay of the first-order satellites within 100 ps, indicating that the non-equilibrium state persists longer than 100 ps. The second-order satellites vanish after about 10 ps: their absence could arise from more diffuse interfaces between the non-thermal and metallic phases [43]. Finally, in the ground state, the first-order and second-order satellites align diagonally and display two maxima along q$_z$ (see Fig. 4a). For the highest measured fluence, only one maximum along q$_z$ is visible at 4 ps and at 49 ps after photoexcitation (see Figs. 4c and Figs. S16-S19), hinting towards a rearrangement of the nano-texture. This is in stark contrast to Sr-doped system without epitaxial strain, which shows a homogenous electronic response[44]. An intuitive interpretation of satellite peak rearrangement is a rotation of interfaces in the nano-texture from being diagonal to being normal to the film surface (see a simulation in Supplementary Figure S21).

Nevertheless, other interpretations are possible, and future modelling and experiments are required to resolve the interface between the equilibrium metallic and the photoinduced phases.

**Conclusion and Outlook**

Our experiments in epitaxially strained $Ca_2RuO_4$ films highlight that photoexcitation above a fluence threshold induces a phase transformation to a long-lived (>100ps) state characterized by (1) an initial sub-ps change in the electronic structure, (2) followed by few-ps lattice expansion, and (3) 10-ps increase of the metallicity. It is plausible that the structural transformation is fast because the low-temperature and high-temperature phases have the same crystal symmetry, which is in contrast to slower, symmetry-changing transitions in vanadates[17,19]. Furthermore, the strain-induced nano-texture may facilitate the structural response; the photoinduced phase may heterogeneously nucleate at inclusions of the incipient long-bond phase stabilized through epitaxial strain. The analysis of the unit cell structure factor in combination with quasistatic first-principles calculations reveals that the long-lived post-transition state is characterized by relatively small amplitude, non-thermal local atomic arrangement, strongly indicating that the transition is not simply driven by heating. The existence of the fluence threshold implies that the material is described by a free energy landscape similar to that shown in Fig. 3d, with two basins of attraction separated by a barrier that requires a sufficiently large fluence to surmount. The slow onset of metallicity observed here provides a counter-example to the generally accepted notion that electronic rearrangements are generically faster than lattice ones. The data and analysis suggest kinetic barriers prevent the electronic state from rapidly equilibrating to the configuration implied by the lattice so that after the initial photoexcitation, the motion along the lattice ($Q_3$) axis of Fig. 3 is more rapid than motion along the electronic ($\phi$) axis.

The experimental results are consistent with the time-dependent first-principles theory and quantitative Landau theory. The comparison to time-dependent many-body theory shows that photoinduced insulator-metal transition originates from long-lived electronic excitations in Mott insulators, which are strongly coupled with lattice degrees of freedom. The slow metallic response originates from kinetic constraints due to the Mott physics: the strong coupling between trapped charge carriers and lattice distortions leads to a complicated evolution within the equilibrium free energy potential from the insulating basin of attraction to the metallic one.

Our work highlights two challenges for ultrafast condensed matter science and some approaches to overcome these challenges. First, the current state of time-dependent theory permits analysis only out to short timescales. A grand challenge is to improve the theory to enable analysis of the large time scales necessary for accessing collective long-wavelength vibrations observed experimentally. The second grand challenge is the difference between the length scales: theory calculates the physics for a local atomic arrangement, including the electronic structure and atomic vibrations. In experiments, long-range collective vibrations and other constraints, such as sample thickness and disorder, are important. Nevertheless, we showed how combining first principles, semi-classical approaches, and quantitative Landau theory elucidates the microscopic mechanisms behind experimentally observed dynamics.

As an outlook, we observed that combined experimental studies of lattice and electronic dynamics provide crucial information enabling the unraveling of dynamically controlled phases in a wide variety of materials. Theoretical challenges include pushing the many-body calculation to longer times, connecting electron-local phonon dynamics to the dynamics of global deformations monitored by Bragg peak intensities, and formulating a nucleation theory of the timescale required for the system to relax from the metastable phase back to the true equilibrium state at very long times. Experimentally, our work further outlines a new pathway to controlling the timescales and nano-textures of photoinduced phase transformations through epitaxial strain. We anticipate it will leverage the interplay between electronic and structural properties in other systems, such as rare-earth nickelates and vanadates [17,45,46].

# Methods

## Thin film synthesis

The film synthesis is identical to Ref.[36]. A $Ca_2RuO_4$ thin film approximately 25 nm thick was grown in a Veeco Gen10 molecular-beam epitaxy (MBE) system on a $(001)_{pc}$-oriented $LaAlO_3$ substrate from CrysTec GmbH. The film was grown at a substrate temperature of 870 °C as measured using a pyrometer operating at 1550 nm. Elemental calcium (99.99% purity) and elemental ruthenium (99.99% purity) were evaporated from a low-temperature effusion cell and an electron beam evaporator, respectively. The films were grown with a calcium flux of $1.8 \times 10^{13}$ atoms·cm$^{-2}$s$^{-1}$ and a ruthenium flux of $1.7 \times 10^{13}$ atoms·cm$^{-2}$s$^{-1}$ in a background of $7 \times 10^{-7}$ Torr of ozone (10% $O_3$ + 90% $O_2$). At the end of the growth, the shutters on both calcium and ruthenium sources were closed, and the sample was cooled down to 250 °C in the same background pressure as used during the growth. All data presented in this paper were collected on exactly the same film.

## X-ray measurements

We have measured the time-resolved X-ray diffraction for a strained $Ca_2RuO_4$ thin film with various pump fluences at the X-ray free-electron lasers at SPring-8 Angstrom Compact free-electron LAser(SACLA), Japan[47]. We have used an IR laser with 1.55 eV photon energy as a pump to excite the system and femtosecond X-ray pulses with 10 keV photon energy as a probe to track the system dynamics. We monitored the out-of-plane 008 Bragg peak with the ground state held at 130 K, below the insulator-metal transition temperature of 200 K. We used cold nitrogen gas flowing from a cryojet to control the temperature in the time-resolved X-ray experiment. The data was collected on an area detector, a multiport charge-coupled device (MPCCD)[48], allowing measuring a slice of the Ewald sphere for a single incident angle. The 3D reciprocal space data was measured by rocking the film with respect to the incident angle $\theta$ with 60 measurements within 1.2 degrees, 23.5°-24.7°. The data in Fig. 1b is the integral of the 3D reciprocal space containing the Bragg peak and the satellite peaks along $q_x$ and $q_y$. The time zero, $\tau = 0$, was set by measuring a test sample and later readjusted by 1.7 ps based on data shown in Fig. 1c: we set the time delay equal to zero to the last data point that is within the experimental uncertainty of the ground state data ($\tau < 0$).

**The time-resolved high-frequency reflectivity measurement**

Optical transient reflectivity measurements were performed using an ultrafast laser supplying pump and probe pulses at 500 kHz. Samples were cooled using a tabletop optical cryostat to a temperature of 131 K. Pump pulses at 1.64 eV (757 nm) provided a fluence of ~0.15 mJ/cm$^2$ and were chopped at 500 Hz with a chopper wheel. Probe pulses at 1.55 eV (800 nm) were used to measure changes in reflectivity with a photodiode using a lock-in amplifier. The transient reflectivity was measured as a function of the time delay between pump and probe pulses which was controlled by an optical delay stage.

**The time-resolved low-frequency reflectivity measurement**

Optical-pump THz-probe (OPTP) measurements were performed in reflection geometry with a 1 kHz Ti:Sapphire amplifier, where 100 fs, 1.55 eV pulses were split into three paths. A portion of the beam was used to generate the THz probe pulse in a ZnTe crystal, and another portion of the beam was used for Electro-Optic-Sampling (EOS) of the THz probe in another ZnTe crystal. The rest of the beam was used as a 1.55eV pump beam. The pump beam impinged on the sample at normal incidence, while the THz probe was incident at approximately 15°. The penetration depth of both beams is larger than the film thickness. The dynamics in Fig. 1c are a 1d THz OPTP scan, where the gate is fixed at the peak of the THz probe and the pump delay is scanned.

**Estimation of the pump-penetration depth**

To estimate the optical absorption spectrum of Ca$_2$RuO$_4$, we used Ref.[49] (see Fig. 3c therein). Because the data only goes up to 5,000 cm$^{-1}$ (1.55 eV = 12,500 cm$^{-1}$), only an order of magnitude estimate can be made. An order-of-magnitude estimate of $\alpha = 10^4$ cm$^{-1}$ extracted from the above reference means that, for a 25 nm film, only 2% of the pump energy is absorbed. Nevertheless, since the estimate has an exponential dependence, it is very sensitive to the actual number used. In an alternative approach, we used the results of Ref.[30] (Fig. 4a therein), which reported the real part of the optical conductivity for 1.55 eV at approximately 600 (Ohm-cm)$^{-1}$. Assuming the imaginary part of the optical conductivity is zero (we couldn't find this quantity reported anywhere in the literature for Ca$_2$RuO$_4$) would give an imaginary index of refraction $\kappa=$ 1.0 and a corresponding absorption coefficient $\alpha = 1.6 \cdot 10^5$ cm$^{-1}$. This latter estimate gives a larger energy absorption of approximately 20%. Regardless of the method, the penetration depth is larger than the film thickness, and we conclude that the film is excited homogeneously.

**Theory**

*Equilibrium free energy:* Here, we follow the analysis in Ref. [8]. Because the lattice is assumed to be harmonic, we may write the result in terms of an electronic order parameter ϕ defined in terms of the difference in the occupancy of the Ru t$_{2g}$-symmetry d orbitals and a normalized lattice displacement Q$_3$ a $F(Q_3, \phi) = \frac{K Q_m^2}{2}\left[\left(\frac{Q_3}{Q_m}\right)^2 - 2\left(\frac{Q_3}{Q_m}\right)\phi\right] + F_{el}$ and $F_{el} = \frac{a}{2}\phi^2 - \frac{b}{3}\phi^3 + \frac{c}{4}\phi^4$, where K is a force constant, and Q$_m$ is the lattice distortion for maximum charge disproportionation ϕ =1. We use the parametrization of the electronic free energy F$_{el}$ determined in Ref. [8] using DFT+DMFT: a=0.7 eV, b=1.7 eV and c=1.28 eV per Ru atom (note the cubic term is allowed because the tetragonal structure of the material means that the xy orbital is different from the xz and yz orbitals so the $\phi = 0$ state is not protected by symmetry). The lattice energy KQ$_m^2$/2 sensitively depends on the substrate and temperature. Ref. [8] argues that in bulk Ca$_2$RuO$_4$, the lattice energy is KQ$_m^2$/2 ≈0.3 eV, while for different substrates it can vary from 0.1 eV<K Q$_m^2$/2 < 0.3 eV. A difference in the impurity solvers used in Ref. [8] and current work leads to a difference in the location of the Mott metal-insulator transition, and we used slightly higher values for the bulk $\frac{K Q_m^2}{2} = 0.35$ eV and the strained film $\frac{K Q_m^2}{2} = 0.30$ eV. We have confirmed that the threshold monotonously decreases with the lattice energy, and the message in Fig. 3c is robust. In the range 0.1 eV<K Q$_m^2$/2 < 0.35 eV, the free energy landscape exhibits two stable local minima with the global minimum corresponding to the insulating phase (ϕ =1) and the local minimum to the metallic phase (ϕ =0). The two phases are separated by a saddle point whose position is almost independent of the value of KQ$_m^2$/2 (see Supplementary Materials). The force constant in the strained case[7] was estimated to K≈34 eV/ Å$^2$ leading to the distortion 0.07 Å < ΔQ$_3$ <0.13 Å, which is very close to the difference of the experimental distortions between the low and high-temperature phase as well as at the maximum excitation strengths (see Fig.2b) if the low-temperature c-axis length is 12.23 Å$^{-1}$.

*Model:* We employed a minimal electron-lattice model for $Ca_2RuO_4$ that includes three $t_{2g}$ orbitals with an average of 4 electrons per site, on-site Slater-Kanamori interactions described below, and an electron-lattice coupling leading to the splitting between $d_{xy}$ and $d_{yz},d_{xz}$ orbitals. The total Hamiltonian is given as H =$H_{kin}$ +$H_{int}$ +$H_{ph}$ +$H_{el\text{-}ph}$. The kinetic energy is given by $H_{kin} = \sum_{k\alpha} h_{k-A\alpha} d^\dagger_{k,\alpha} d_{k,\alpha} + \vec{E}\vec{D} \sum_k [d^\dagger_{k,dxy} d_{k,dyz} + d^\dagger_{k,dxy} d_{k,dxy} + h.c.]$, where $d_{k\alpha}$ are the annihilation operators for orbitals $\alpha \in (d_{xy},d_{yz},d_{xz})$ and momentum k, and A is the vector potential. The kinetic matrix elements are given by $h_k$ and for practical reasons we will solve the problem on the Bethe lattice with a semicircular self-consistency. We match the bandwidth of bands as $W_{dyz} \approx W_{dxz} \approx 1.0$ eV and $W_{dxy} \approx 2.0$ eV to be consistent with the ARPES data in Ref. [6]. The chemical potential μ is chosen such that the system is at half-filling. We excite the system with a short electric pulse $E(t) = -\partial_t A = E_0 \sin[\Omega(t-t_0)] e^{-4.6(t-t_0)^2/t_0^2}$, where the width of the pulse $t_0$ is chosen such that the pulse includes one oscillation, and we have adjusted the amplitude $E_0$ to fix the amount of double occupancy. While the direct optical transitions are not allowed within the d-orbital multiplet, the small distortion of the octahedra leads to small but finite dipolar matrix elements D=0.4 and allows for a direct charge-transfer excitation. We model the interacting part within the Slater-Kanamori representation

$$H_{int} = U \sum_{i,\alpha} n_{i,\alpha\uparrow} n_{i,\alpha\downarrow} + \sum_{i,\alpha<\beta} \sum_{\sigma,\sigma'} (U' - J_H) n_{i,\alpha\sigma} n_{i,\beta\sigma'} + J_H \sum_{i,\alpha<\beta} d^\dagger_{i,\alpha\uparrow} d^\dagger_{i,\alpha\downarrow} d_{i,\beta\uparrow} d_{i,\beta\downarrow} + d^\dagger_{i,\alpha\uparrow} d^\dagger_{i,\beta\downarrow} d_{i,\alpha\downarrow} d_{i,\beta\uparrow}$$

and use the established values of the U = 2.3 eV and $J_H$ = 0.4 eV[6].

First-principle calculations have established that the difference in the change occupation is strongly coupled to a lattice mode[8,23,24] parameterized as $Q_3 = \left(\frac{1}{\sqrt{6}}\right) \cdot (2c - a_0 - b_0)$. The lattice part of the energy is given by $H_{e-ph} = KQ_m^2 \left[\left(\frac{Q_3}{Q_m}\right)^2 - 2\frac{Q_3}{Q_m}\varphi\right] + \frac{KQ_m^2}{\omega_0^2}\left(\frac{Q_3}{Q_m}\right)^2$, where $\omega_0$ is the phonon frequency. The precise parameters depend on the coupling of this mode to the other modes in the solid and on the substrate strain.

*Method:* We solve the problem within the time-dependent DMFT and explicitly simulate the time-evolution during and after the excitation using the numerical library Nessi[27]. We use a multiband extension of non-crossing approximation (NCA) as an impurity solver (see Ref. [50] for details). To reach longer final times, we solved the problem on the Bethe lattice and consider only intraband hopping leading to the simplified diagonal DMFT self-consistency. Moreover, we have explicitly checked the effect of the magnetic ordering by considering the photodoping of the anti-ferromagnetic solution. The final self-consistency is given by $\Delta_{\alpha,\uparrow}(t, t') = (W_\alpha/4)^2 \, G_{\alpha,\downarrow}(t, t')$, where $(G_{\alpha\uparrow})(\Delta_{\alpha\uparrow})$ is the local, orbital and spin-dependent Greens (hybridization) function defined on the Keldysh contour. We treat the electron-phonon problem on the mean-field level leading to the equations of motion $\dot{Q}_3 = \omega_0 \Pi$, $\dot{\Pi} = -\omega_0 Q_3 + g\phi$, where $g = \sqrt{(2\omega_0^2 \, K \, Q_m^2)}$. The stationarity of these equations gives us the linear dependence of the lattice distortion versus the order parameter orbitals $Q_3 = g\,\phi/\omega_0$. In simulations, we have used lattice frequency to $\omega_0/W_{dxy} = 1/8$ and the lattice energy $\frac{KQ_m^2}{2} = 0.35$ eV to simulate the bulk and $\frac{KQ_m^2}{2} = 0.3$ eV for strained thin films. The scaling to lower lattice frequencies is performed in the Supplemental Materials (see Fig. S10).

**Acknowledgments**

The work was primarily supported by U.S. Department of Energy, Office of Science, Office of Basic Energy Sciences, under Contracts No. DE-SC0019414 (X-ray experiments and interpretation A.V., O.G., J.R., K.M.S., A.S.; thin film synthesis: H.N., N.J.S.; DFT calculations: J.Z.K, G.K, N.A.B, high-frequency reflectivity: R.R., J.W.H.). AJM acknowledges support from the U.S. Department of Energy, Office of Science, Office of Advanced Scientific Computing Research, Scientific Discovery through Advanced Computing (SciDAC) program, under Award No. DE-SC0022088. The XFEL experiments were performed at the BL3 of SACLA with the approval of the Japan Synchrotron Radiation Research Institute (JASRI) (Proposal No. 2019A8084). D.G. is supported by Slovenian Research Agency (ARRS) under Program J1-2455 and P1-0044. K.K., V.R. and R.D.A supported under NSF DMR-1810310. The calculations have been performed using a software library developed by M. Eckstein and H.U.R. Strand. The Flatiron Institute is a division of the Simons Foundation. We thank Ben Gregory and Ziming Shao for measuring the resistivity.

**Figures**

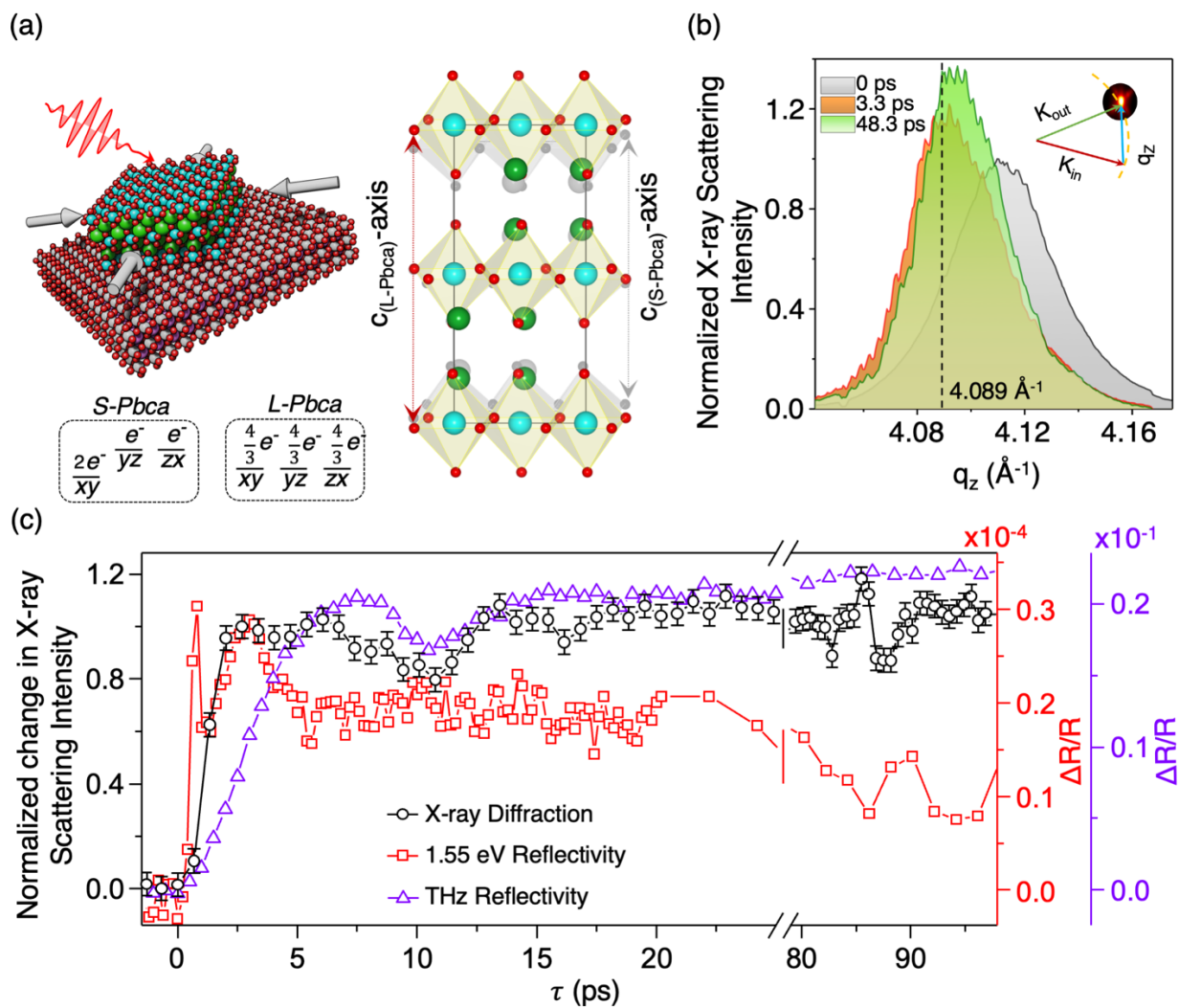

**Fig. 1| Photoinduced structural change and insulator-to-metal transition.** (a) Schematic representation of the epitaxially strained thin-film (O-red, Ca-green, Ru-cyan, La-magenta, Al-gray), structural phase transformation from L-Pbca (color) and S-Pbca (shaded), and electronic configuration of Ru d-orbitals in $Ca_2RuO_4$. (b) Photoinduced dynamics of 008 Bragg peak of strained $Ca_2RuO_4$ thin film at pump fluence of 50 mJcm$^{-2}$, where the peak-shift towards a lower momentum transfer $q_z$ within 3.3 ps indicates a lattice expansion. (c) The time-resolved normalized change in the scattering intensity (black squares, incident pump-fluence 50 mJcm$^{-2}$) at a fixed wavevector, $q_z = 4.089$ Å$^{-1}$, increases in about 2.5 ps and persists for $\tau \leq 100$ ps. The time-resolved high-frequency reflectivity (red circles, E=1.55 eV, incident pump-fluence 0.14 mJcm$^{-2}$) increases rapidly, within 1 ps, shows a peak coincident with the lattice expansion, and decays slowly within 100 ps. The time-resolved low-frequency reflectivity (purple triangles, THz-bandwidth from 0.8 meV to 10 meV, incident pump-fluence 15.1 mJcm$^{-2}$) signal increases within about 8 ps and persists for 100 ps. The time-resolved x-ray data and low-frequency reflectivity were measured after photoexcitation

(pump) with a E=1.55 eV femtosecond laser, the time-resolved high-frequency reflectivity with a E=1.64 eV femtosecond laser.

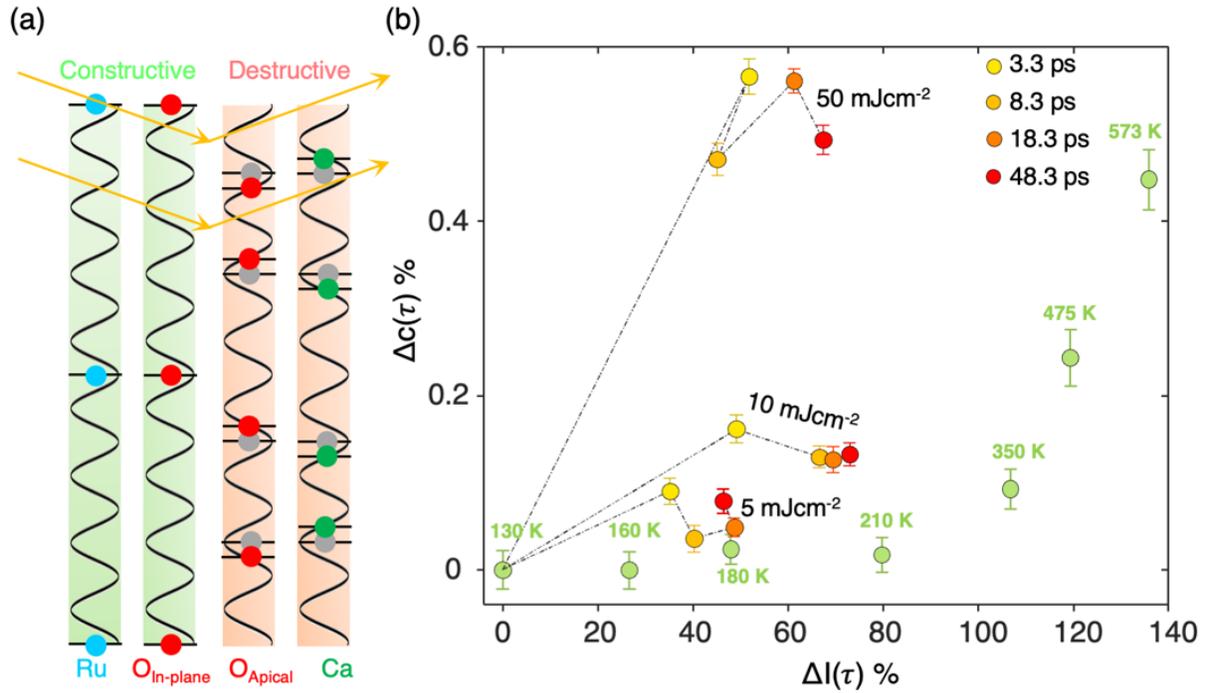

**Fig. 2| Photoinduced structural rearrangements of the Ca$_2$RuO$_4$ unit cell.** (a) The schematic diagram for the Bragg-plane shifts from S-Pbca (shaded colors in the background) to L-Pbca (bold colors in the foreground) shown with the x-ray wave for the 008 Bragg condition (eight wavelengths $\lambda$ within a unit cell). The positions of different atomic planes are separated horizontally into four groups, and the shifts are enlarged for clarity. The Ru and in-plane O atoms scatter coherently (separated by $n\lambda$, $n$=integer). In S-Pbca, the Ca and apical O are close to the destructive interference condition (separated by $(n+\frac{1}{2})\lambda$). Transition to L-Pbca moves the structure away from destructive interference and increases the structure factor. (b) The photoinduced change of the c-lattice parameter, $\Delta c(\tau)$, as a function of the relative change in integral intensity, $\Delta I_{008}(\tau)$, for multiple measured fluences (5 mJcm$^{-2}$, 10 mJcm$^{-2}$, 50 mJcm$^{-2}$). For comparison, the quasi-static data are also shown (green circles).

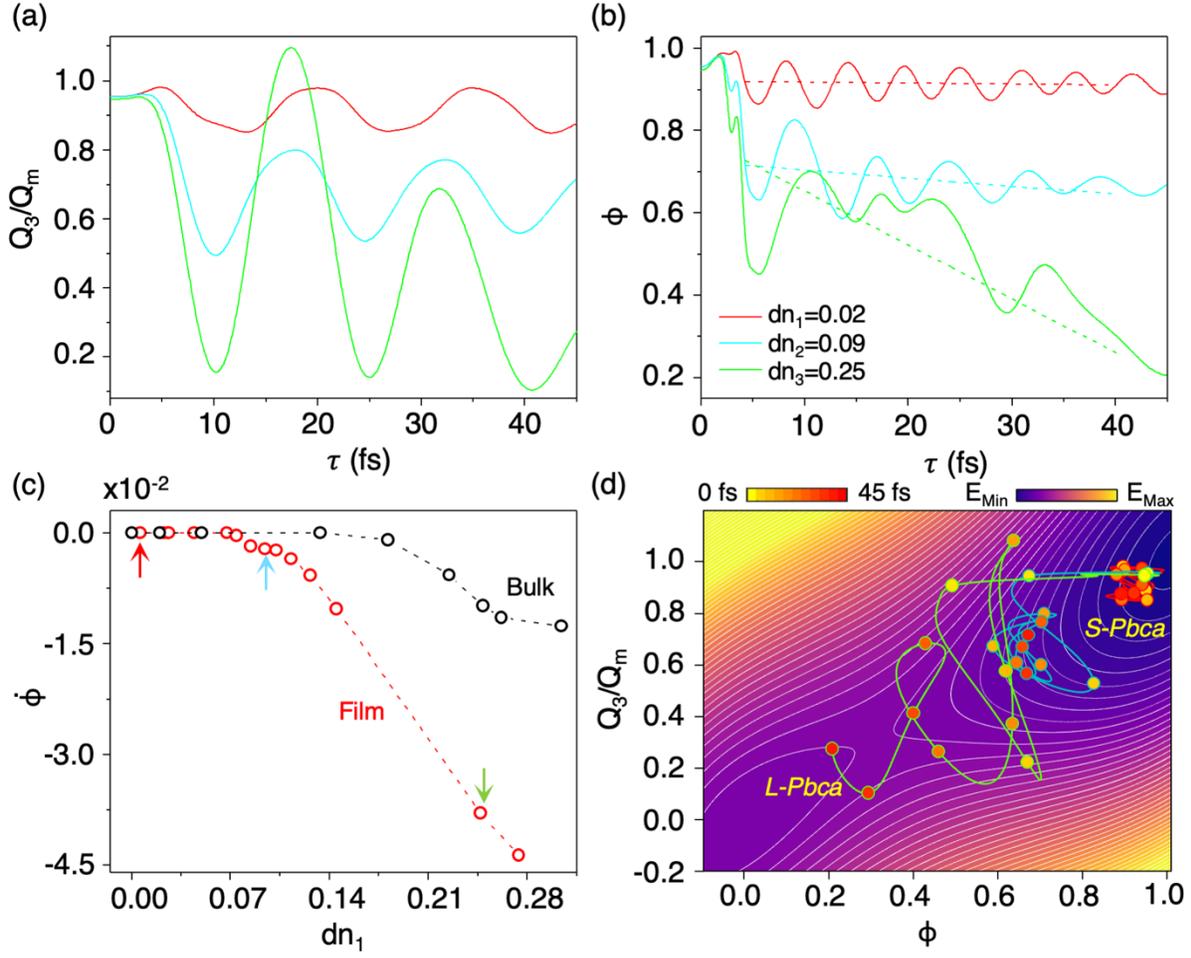

**Fig. 3| Results from time-dependent dynamical mean-field theory and quantitative Landau theory.** Short time evolution of the artificial high-frequency lattice mode $Q_3$ (a) and charge disproportionation $\phi = n_{xy} - (n_{xz} + n_{yz})/2$ (b) for increasing photodoping $dn_1$ (in percent of charge carriers per unit cell). (c) The slope of the long-time dynamics, $\dot{\phi}$, in charge disproportionation (fit marked with dashed lines in (b)) versus the photodoping $dn_1$ for the lattice energies corresponding to bulk $[\frac{K Q_m^2}{2} = 0.35\ eV]$ and thin films $[\frac{K Q_m^2}{2} = 0.30\ eV]$. (d) The trajectory of order parameters (excitations marked with arrows in (c)) embedded in the equilibrium Landau-Ginzburg potential extracted from Ref.[8].

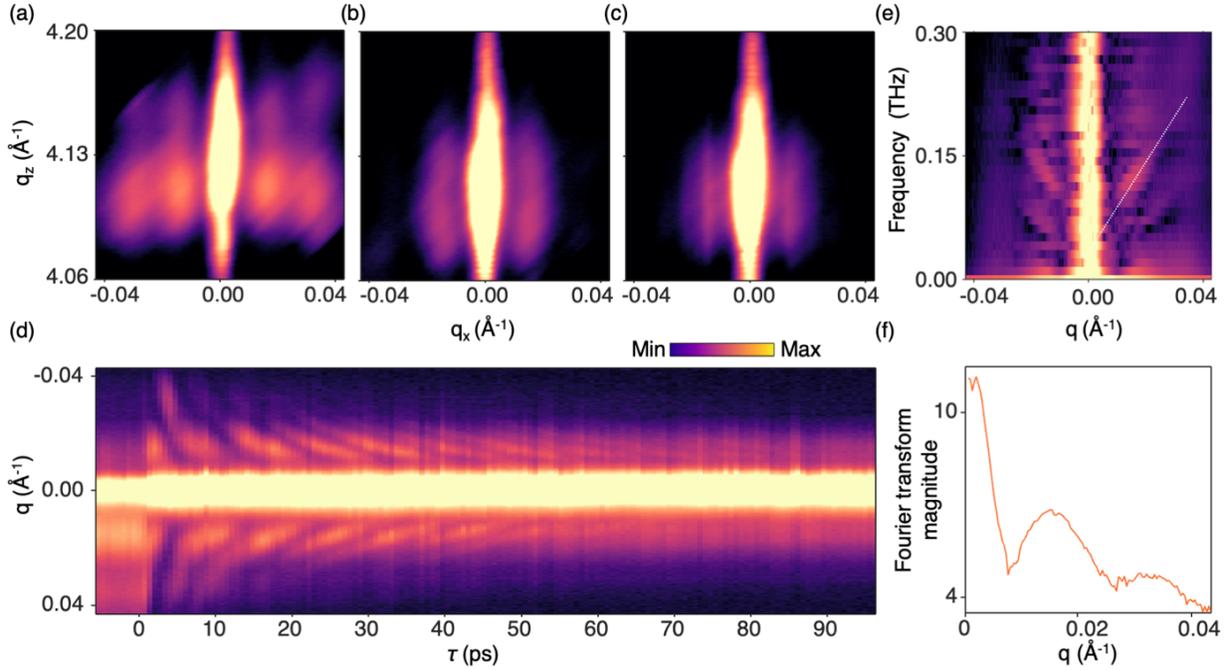

**Fig. 4| Dynamics in the nano-texture in the strained $Ca_2RuO_4$ thin film.** (a-c) Measured 3D time-resolved X-ray scattering shown on a $q_x$-$q_z$ plane passing through the 008 Bragg peak for a pump fluence of 50 mJcm$^{-2}$ in the ground state (a), at $\tau$=3.3 ps (b), and at $\tau$=48.3 ps (c) (logarithmic scale, $q_x$ along **a**=[100], $q_y$ along **b**=[010], and $q_z$ along **c**=[001] in the orthorhombic Pbca notation). Satellite peaks along the $q_x$ direction signify a periodic nano-texture along [100] (**a**-axis, oriented along the film surface). In the ground state, the periodicity of the nano-texture is $\Lambda_{[100]} = 2\pi/\Delta q_x \approx 40$ nm, where $\Delta q_x \approx 0.015 \text{Å}^{-1}$ is the position of the first-order satellite peak. (d) Time-resolved X-ray scattering with a higher time resolution than (a-c), measured on a single slice of the Ewald sphere and projected onto $q_x$. (e) Frequency dispersion relation extracted as the magnitude of the discrete Fourier Transform of the data shown in (d). (f) A line intensity profile along the dotted white line in Fig. 4e.

**Supplementary Material**

**S1. The relation of lattice expansion with coherent longitudinal acoustic phonons**

In our study, photoexcitation rapidly changes electronic orbital occupancies. Because the films are thinner than the optical penetration depth, the change may be considered homogeneous throughout the film thickness. The new electronic configuration creates a new displaced potential to which the lattice responds by launching an acoustic wave. For a free uniformly expanded sphere, the resulting modes were calculated by Lamb in 1881[1], who found a breathing mode with a frequency set by the sound velocity and sphere radius. In our system, the acoustic wave propagates from the free surface to the interface with the substrate, where some fraction of the wave is reflected back into the film and some fraction propagates into the substrate. In the experimental situation, the wave is also damped by decay into other modes and interface roughness, not included in the calculation. We modeled the situation using a one-dimensional nearest neighbor force constant model with a relatively thin film coupled to a semi-infinite substrate with different force constants in film and substrate and an initial condition corresponding to a uniformly strained film but no strain in the substrate. Representative results are shown in Fig. S2. Panel (a) shows the normalized displacement of selected layers as a function of time. At time $\tau = 0$ one sees a compressive uniform strain, with the atom at the vacuum interface having the largest displacement (normalized to -1) and successive layers having proportionately smaller displacements consistent with a uniform strain. The time evolution of the curves shows the strain relaxation propagating from the free surface, being partly reflected at the interface, and oscillating in a decaying manner. Panel (b) shows the corresponding Bragg peak, obtained by Fourier transforming the positions of the atoms in the film layer only. The broadening of the peak is a consequence of the finite thickness of the film. A relaxation of the peak from the initial value corresponding to the strained lattice towards the final value corresponding to the relaxed lattice is evident, along with small oscillations of the peak around its final position.

We measured the diffraction intensity at a fixed $q_z$ on the tail of the peak such that the peak center does not move past the fixed $q_z$ (see peak shapes and the dashed line in Fig. 1b). Therefore, the change of the lattice constant is monotonically mapped on the diffracted intensity. The distance between time zero and the first local extremum equals half of the period (see Fig. S2, a). From the data in Fig. 1c, we estimate the oscillation period to be $\tau = 1/\nu \approx 8$ ps, which is consistent with a calculated period [2] $\tau = v_c/(2d) = 7$ ps, where $v_c = 7$ nm/ps is the speed of sound and $d =$ 25 nm is the film thickness. The speed of sound was calculated for bulk $Ca_2RuO_4$ with DFT (see S2). The expected oscillations visible in Fig. S2 are buried within the experimental uncertainties and not visible in Fig. 1c, suggesting the oscillations of the lattice constant are damped rapidly by scattering into other modes not included in the calculation.

**S2. Density functional theory calculations for elastic constants and estimating the structure factor**

Structural relaxation of strained $Ca_2RuO_4$ was calculated via density functional theory (DFT) in the local density approximation with Hubbard U[3] - (LDA+U) as implemented in VASP.6.2.0[4-6]. The electrons included in the valence of the projector augmented wave potentials were: $3s^23p^64s^2$ for Ca, $4s^24p^64d^75s^1$ for Ru, and $2s^22p^4$ for O. We use a tetragonal cell containing 4 formula units to describe both ferromagnetic and antiferromagnetic Ru configurations, where U=2.3 eV and J=0.35 eV as found in Ref.[7]. The structure and elastic constants were converged to within 0.1% for the lattice constants and 2% for the elastic constants for a 550 eV plane-wave cutoff and Monkhorst–Pack grid of 10×10×6 k-points with a force convergence tolerance of $10^{-3}$ eV/Å , when compared to a 700 eV plane-wave cutoff, in the bulk structure. The elastic constants were calculated using VASP's built-in scheme as well as the procedure outlined in Ref. [8](on a mesh in increments of 0.05% up to ±0.5%, while relaxing the internal coordinates). The results between the two methods agree within ±3%. The calculated elastic constants from the two methods are included in Table 2. The speed of sound, $v_c$, (along c) is estimated by[9, 10]:

$$v_c^2 = \frac{(c_{55} + c_{33}) + \sqrt{(c_{55} + c_{33})^2 - 4c_{55}c_{33}}}{2\rho}$$

where $\rho$ is the density (2.88 AMU/Å$^3$ from LDA+U), and $c_{ij}$ are the elastic constants. The resulting speed of sound is 7.3 nm/ps using coefficients from VASP (7.2 nm/ps using Ref.[8]).

| Method | $c_{11}$ | $c_{22}$ | $c_{33}$ | $c_{44}$ | $c_{55}$ | $c_{66}$ | $c_{12}$ | $c_{13}$ | $c_{23}$ |
|---|---|---|---|---|---|---|---|---|---|
| VASP | 1494 | 2039 | 2546 | 573 | 678 | 899 | 1379 | 1215 | 1333 |
| Ref.[8] | 1505 | 2031 | 2479 | 583 | 671 | 875 | 1385 | 1194 | 1302 |

**Table 1.** DFT-calculated elastic constants for bulk $Ca_2RuO_4$, in units of kBar.

In exploring the sensitivity of the 008 Bragg peak to structural distortions, five structures were compared: bulk antiferromagnetic (AFM) and ferromagnetic (FM) configurations (labeled b-AFM and b-FM), AFM and FM configurations strained to the in-plane lattice parameters of LaAlO$_3$ (labeled s-AFM and s-FM), and a lower energy s-AFM structure (labeled s-AFM(P2$_1$/c)) described in what follows. In LDA+U, the LaAlO$_3$-strained AFM structure (Pbca, #61 – D$_{2h}$ point group) is dynamically unstable with two phonon instabilities ($\Gamma_2^+$(B$_{1g}$) and $\Gamma_4^+$(B$_{2g}$) both primarily of O character) leading to P2$_1$/c (#14 – C$_{2h}$). We find that the $\Gamma_2^+$(B$_{1g}$) instability lowers the energy by 9.19 meV/f.u as compared to 2.52 meV/f.u. for the $\Gamma_4^+$(B$_{2g}$). The P2$_1$/c structure arrived at from the $\Gamma_2^+$(B$_{1g}$) instability is labeled s-AFM(P2$_1$/c). The relaxed lattice constants of these different structures are presented in Table 2.

In each of these structures, symmetry-conserving distortions were applied to model the 008 Bragg peak intensity. X-ray diffraction Bragg peak intensity was calculated using the XrayUtilities software[11]. Changes in the 008 Bragg peak are only sensitive to constructive/destructive interference induced by the motion of atoms along the c-axis. We first explored how fixing the c-axis to different lengths affected the 008 Bragg peak; the results are presented in Figure S3. In addition to changes in the c-axis, we explored the motion of the internal degrees of freedom by using a basis of symmetry-adapted modes describing the motion of atoms along the c-axis that preserve the space group symmetry as found in the ISOTROPY Software Suite[12, 13]. These fully symmetric modes are labeled $\Gamma_1^+$(Ca), $\Gamma_1^+$(O$_{Ru}$), and $\Gamma_1^+$(O$_{Ca}$) to reference the Ca, O in the Ru-O$_2$ plane (O$_{Ru}$), and the O in the Ca-O plane (O$_{Ca}$). The results are presented in Figures S4-S6 for different modes.

| Structure | Lattice constants | | |
|---|---|---|---|
| | a (Å) | b (Å) | c (Å) |
| b-AFM | 5.224 | 5.651 | 11.535 |
| b-FM | 5.283 | 5.378 | 11.858 |
| s-AFM | 5.332 | 5.332 | 11.889 |
| s-FM | 5.332 | 5.332 | 11.856 |
| s-AFM(P2$_1$/c) | 5.332 | 5.332 | 11.842 |

**Table 1.** DFT-relaxed lattice constants for the structures explored in this work. Note that the in-plane lattice constants (a and b) of the strained structures are fixed to the DFT values of LaAlO$_3$ that are expected for coherent epitaxial strain. We note that the lattice parameters found in DFT underestimate the experimental lattice parameters, as expected, due to under-bonding in the local density approximation.

## S3. Scaling with decreasing phonon frequency

In the theoretical analysis of the main text, we have considered a phonon with artificially increased lattice frequency. Here, we will show that scaling to lower phonon frequencies leads to qualitatively similar, but slower, dynamics with the photo-induced drift toward a metallic solution, see Fig. 3. In all calculations, we have fixed the dimensionless electron-lattice interaction $\lambda = g^2/(t\,\omega_0) = 0.72$ and modified the frequency. In Figure S10, we show the evolution of the lattice mode $Q_3$ and the charge order parameter $\phi$ for the reduced frequency of the lattice mode $\omega_0$=0.125 eV. The evolution of the lattice mode $Q_3$ shows a clear doubling of the period of oscillation for the weakest excitation $dn_1$=2.4 %. For stronger excitations, the initial drop of the lattice distortion is delayed, but the longer time evolution gets nontrivial.

The evolution of the electronic order parameter exhibits a short-time dynamic which is almost equivalent for the fast and slow lattice modes. This points to the electronic origin of the short-time dynamics, namely the photo-induced charge-transfer excitation, and is consistent with the short-time decoupling between electronic and lattice degrees of freedom in the experiment (compare the X-ray response (black symbols) with the high-frequency response (red symbols) in Figure 1c). Subsequent oscillations show the expected slowdown in the dynamics of the slow mode and the delay in the long-time drift. The dynamics are, therefore, qualitatively similar to the ones presented in the main text but with obviously slower dynamics due to the reduced frequency of the lattice mode. We can propose that understanding the long-time limit of this problem poses a significant challenge for the development of future theoretical methods based on either time-truncation or compression techniques [14, 15].

## S4. Free energy landscape analysis

The free energy landscape depends sensitively on the size of the lattice energy $KQ_m^2/2$ and in Ref.[16], it was estimated that it can vary in the range 0.1 eV<$KQ_m^2/2$<0.3 eV. The landscape for various lattice energies is presented in Fig. S11, where we can see that the insulating phase $\phi = 1$ is always the ground state, and the metallic state $\phi = 0$ is a local minimum. The position of the saddle point only weakly depends on the lattice energy $KQ_m^2/2$ and is centered around $\phi = 0.5$.

## S5. Three-dimensional reciprocal space data collection

We collected 3D reciprocal space maps by rocking the crystal and recording a diffraction pattern at each rocking angle. For each dataset shown in Fig. 4a-c and Figs. S13-S14, we collected 61 measurements in $\theta$ from 23.5-24.7 degrees, with 0.02 step size. A typical diffraction image is shown in Fig. S22. Each scan took about 6 minutes of acquisition time. The data was subsequently transformed into a cartesian coordinate system in reciprocal space, which is shown in the paper. The position of the Bragg spot deviated slightly from its expected position in the scattering plane. We used no monochromator and anticipate that the deviation was present because of the angular jitter of the FEL. We corrected the slight deviation in post-analysis.

## Supplementary Figures

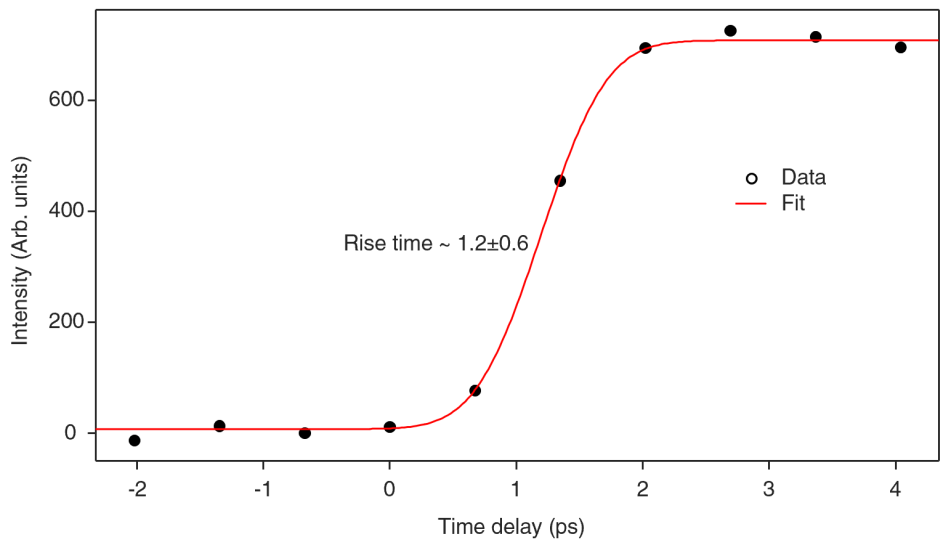

**Figure S1. Estimating the rise time from data shown in Figure 1c of the main test.** The time-resolved normalized scattering intensity of 008 Bragg shown in Figure 1 b of the main text fitted to determine a rise time of 1.2±0.6 ps.

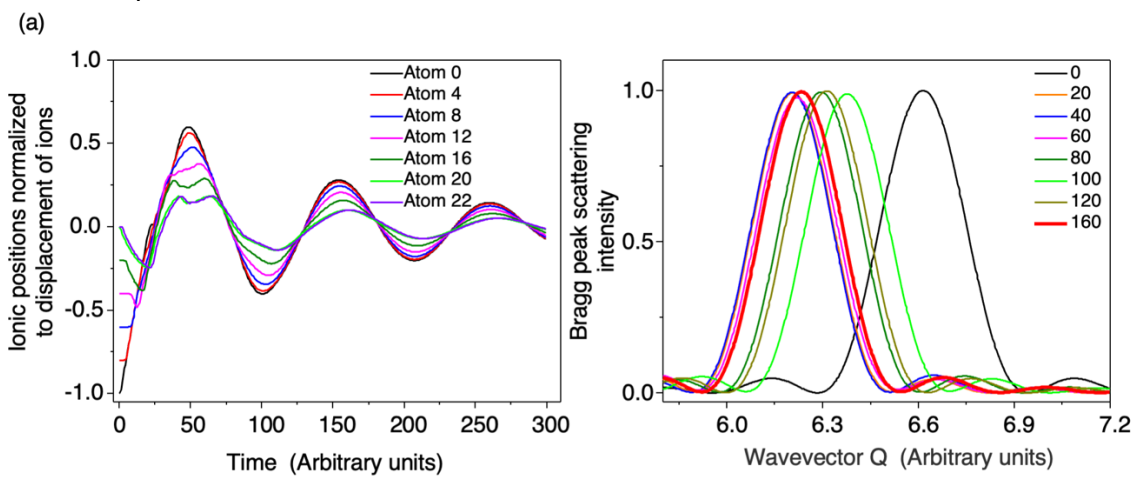

**Figure S2.** (a) Ionic positions normalized to the displacement of the ion at the vacuum interface plotted against time for ions at layers 0 (vacuum interface), 4, 8, 12, 16, 20 (adjacent to the substrate) calculated for a 21-unit cell thick film on a substrate chosen large enough (240 layers) to be effectively semi-infinite, and with a substrate sound velocity 8% less than the film sound velocity (atom 22 is in the substrate). (b) Bragg peak scattering intensity computed from time-dependent atomic positions, showing the evolution from a value of 6.6 (strained lattice) to a value of $6.28 = 2\pi$ (relaxed lattice). The thick red curve is approximately the long-time position. Numbers in the legend indicate the time sequence from (a).

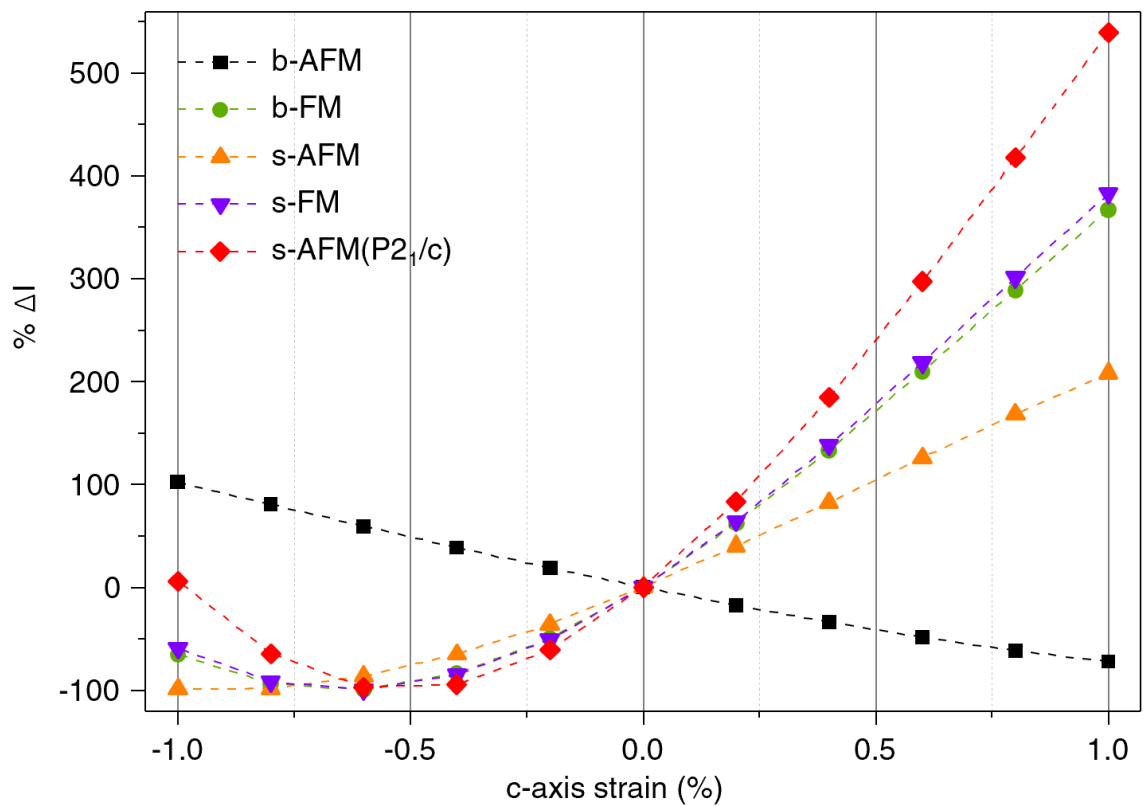

**Figure S3.** Change in the 008 Bragg peak intensity as a function of c-axis strain for bulk and biaxially strained $Ca_2RuO_4$ simulated via DFT, allowing for relaxation of the internal atomic coordinates.

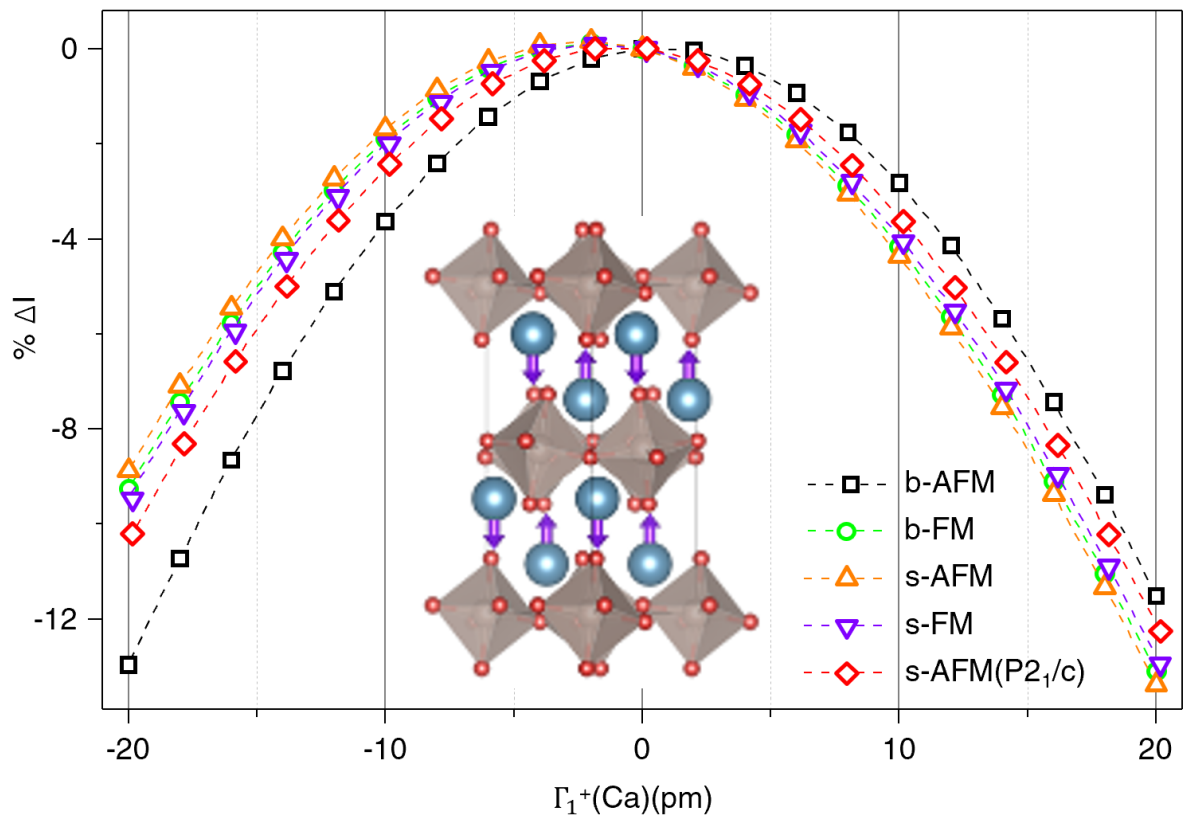

**Figure S4.** Change in the 008 Bragg peak intensity as a function of symmetry-constrained Ca displacements, $\Gamma_1^+(Ca)$, for bulk and strained $Ca_2RuO_4$ simulated via DFT. The inset shows the $\Gamma_1^+(Ca)$ mode.

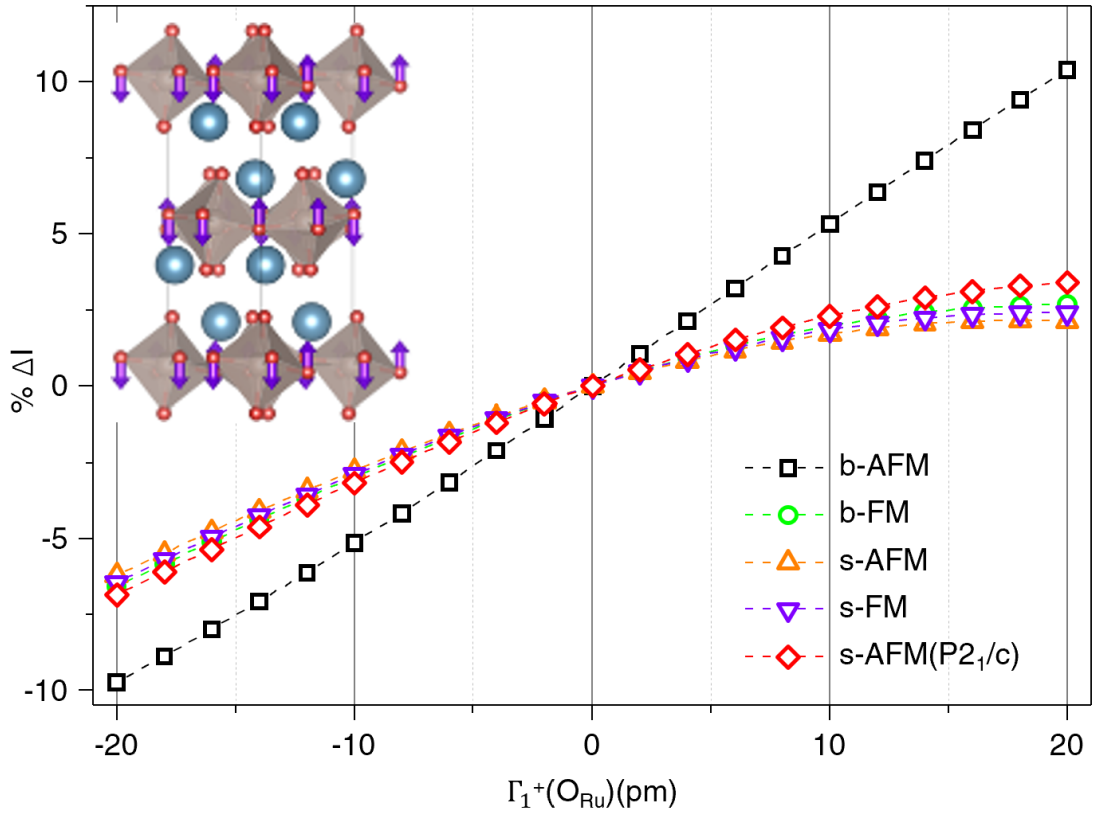

**Figure S5.** Change in the 008 Bragg peak intensity as a function of symmetry constrained O displacements within the Ru-O$_2$ plane, $\Gamma_1^+(O_{Ru})$, for bulk and strained Ca$_2$RuO$_4$ simulated via DFT. The inset shows the $\Gamma_1^+(O_{Ru})$ mode.

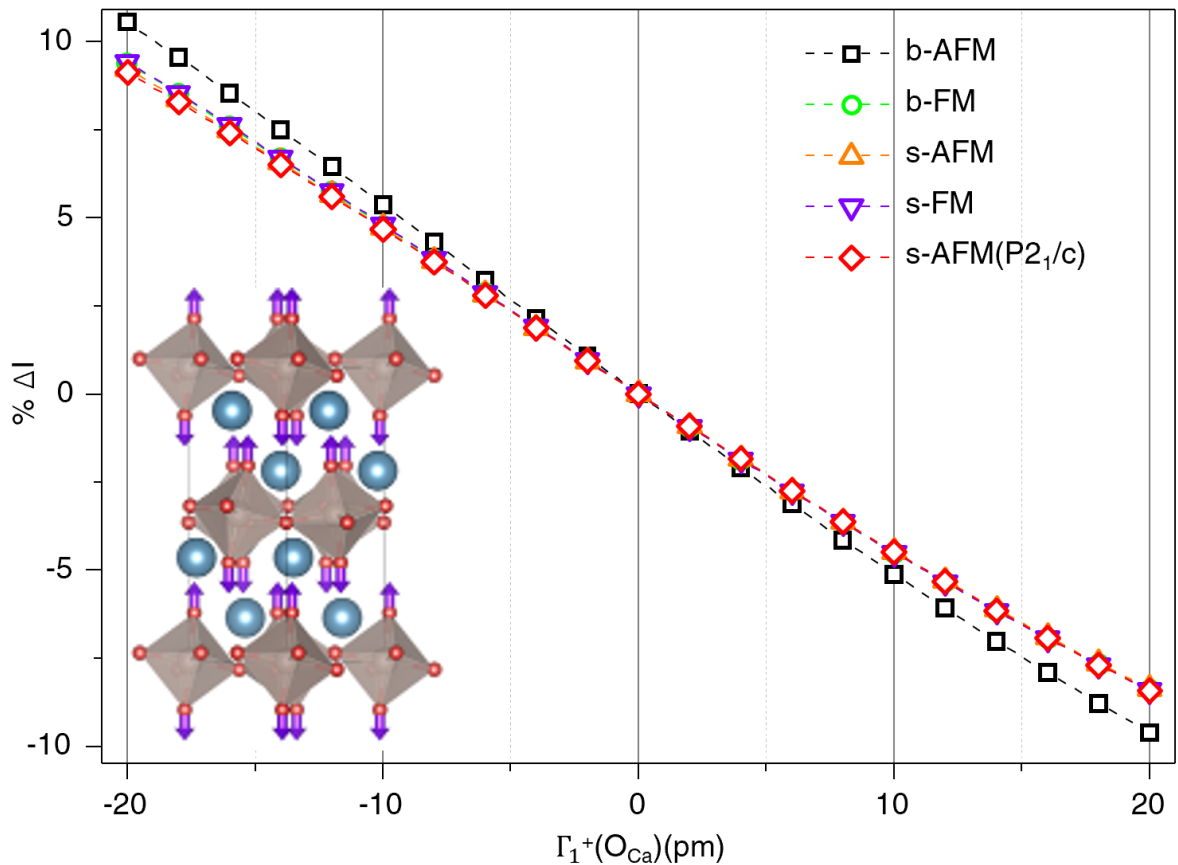

**Figure S6.** Change in the 008 Bragg peak intensity as a function of symmetry-constrained O displacements within the Ca-O2 plane, $\Gamma_1^+(O_{Ca})$, for bulk and strained $Ca_2RuO_4$ simulated via DFT. The inset shows the $\Gamma_1^+(O_{Ca})$ mode.

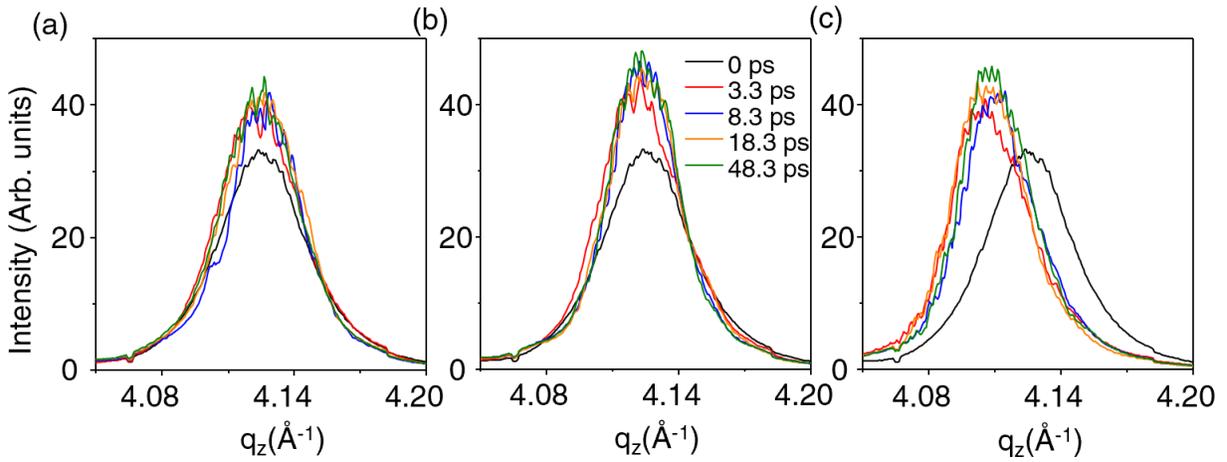

**Figure S7. Photoinduced dynamics of 008 Bragg peak of strained $Ca_2RuO_4$ thin film at different fluences.** The films were excited with (a) 5 mJcm$^{-2}$, (b) 10 mJcm$^{-2}$, and (c) 50 mJcm$^{-2}$. After photoexcitation at lower fluences (a,b), the c-lattice spacing remains mostly unchanged. At the highest fluence, the peak shifts towards a lower momentum transfer $q_z$ within 3.3 ps, indicating a lattice expansion.

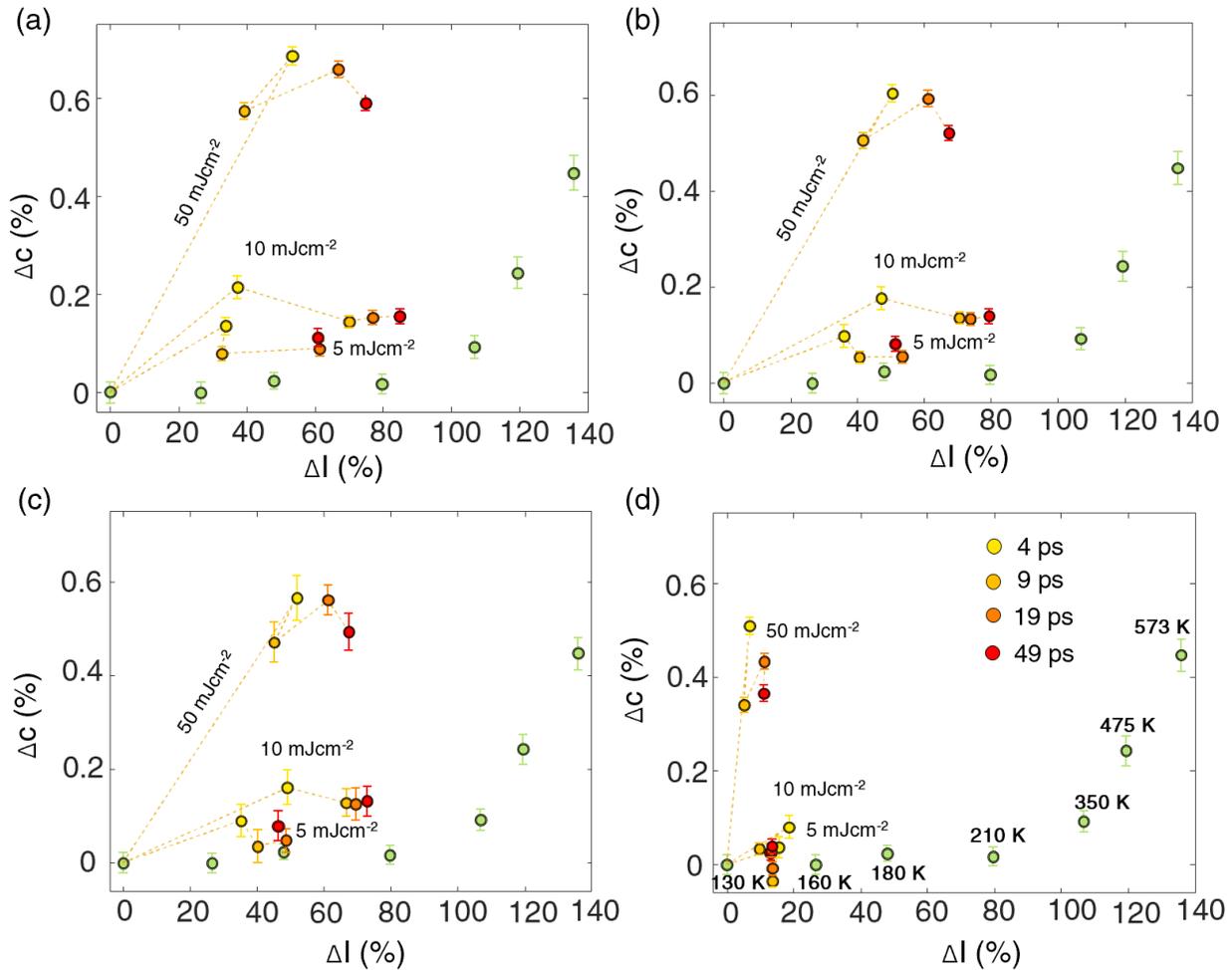

**Figure S8. Figure 2b from the main text reproduced with different widths for integration around the Bragg rod.** The three-dimensional reciprocal space data at XFEL was averaged in a square rod around the Bragg rod with (a) full reciprocal space (rod width 0.08Å$^{-1}$ x 0.08Å$^{-1}$) including satellite peaks, (b) a rod 0.0042Å$^{-1}$ x 0.0042 Å$^{-1}$ pixels wide (shown in the main text), (c) 0.0021Å$^{-1}$ x 0.0021 Å$^{-1}$ wide, and (d) 0.0011Å$^{-1}$ x 0.0011 Å$^{-1}$ wide. In the equilibrium data (green circles), collected with the Rigaku Smartlab diffractometer, satellite fringes are not detectable. Therefore, for a comparison between SACLA and lab-XRD data, in the main text, we averaged over 20 pixels in the reciprocal space, which captures the Bragg rod but does not include the satellite peaks. Even with the satellite peaks in (a), the integral intensity does not reach the change in the integral intensity of the quasi-static data.

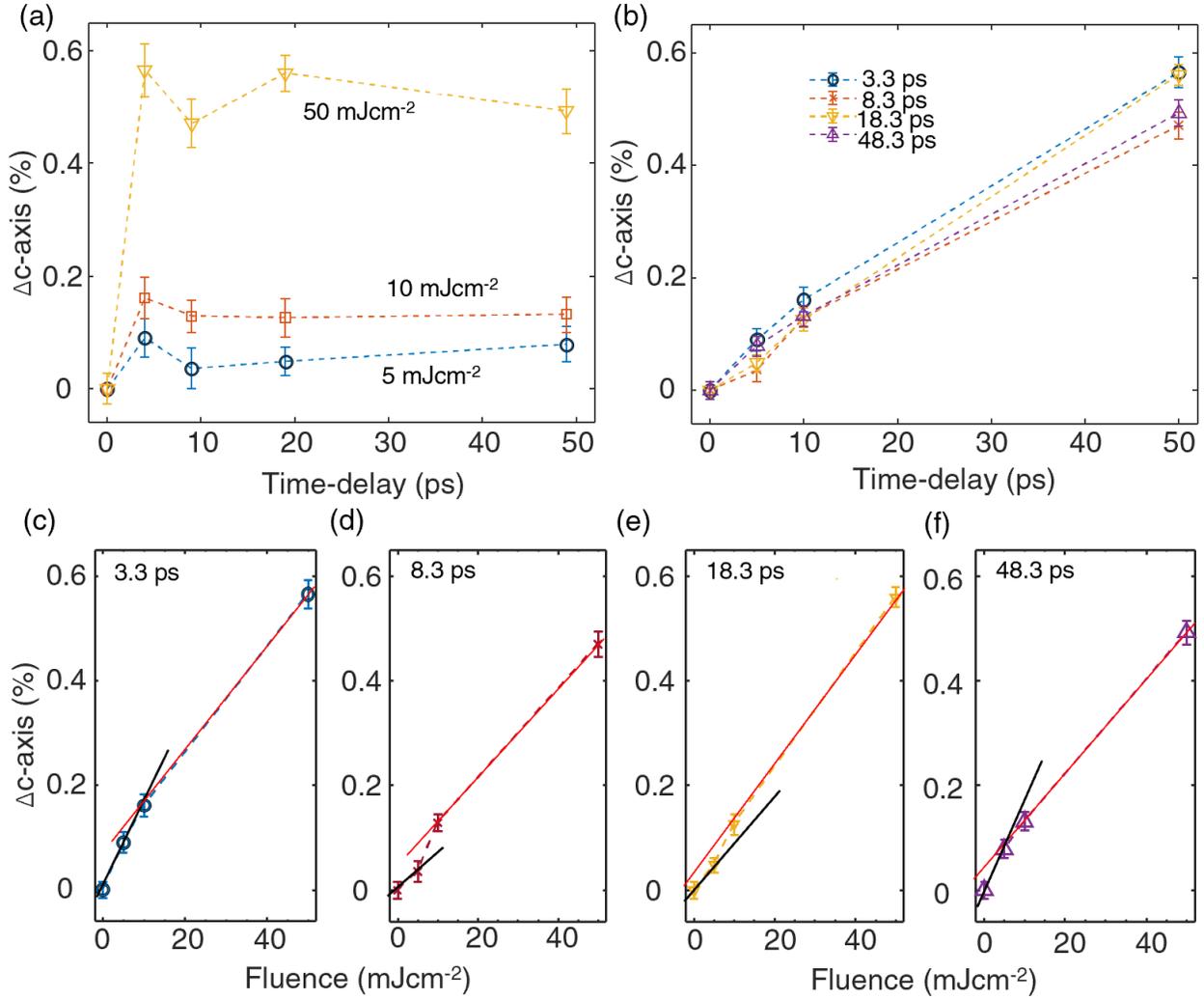

**Figure S9. Fluence-dependent Ca$_2$RuO$_4$ lattice expansion.** (a) The c-lattice spacing (inversely proportional to the position of the 008 Bragg peak) as a function of the time delay was extracted by fitting Gaussian functions to data shown in Fig. S7. (b) The c-lattice spacing as a function of fluence for different time delays. (c-f). Nonlinear increase in c-axis with fluence at fixed time delays. The blue line shows the slope determined from the first two points: 0 mJcm$^{-2}$ (ground state) to 5 mJcm$^{-2}$. The red line shows the slope determined from the last two points: 10 mJcm$^{-2}$ and 50 mJcm$^{-2}$. For small time delays, the c-lattice constant is potentially affected by lattice oscillations. A fluence threshold is apparent for 48.3 ps, where the slopes of the black and red lines differ, and the lattice vibrations are largely damped.
36

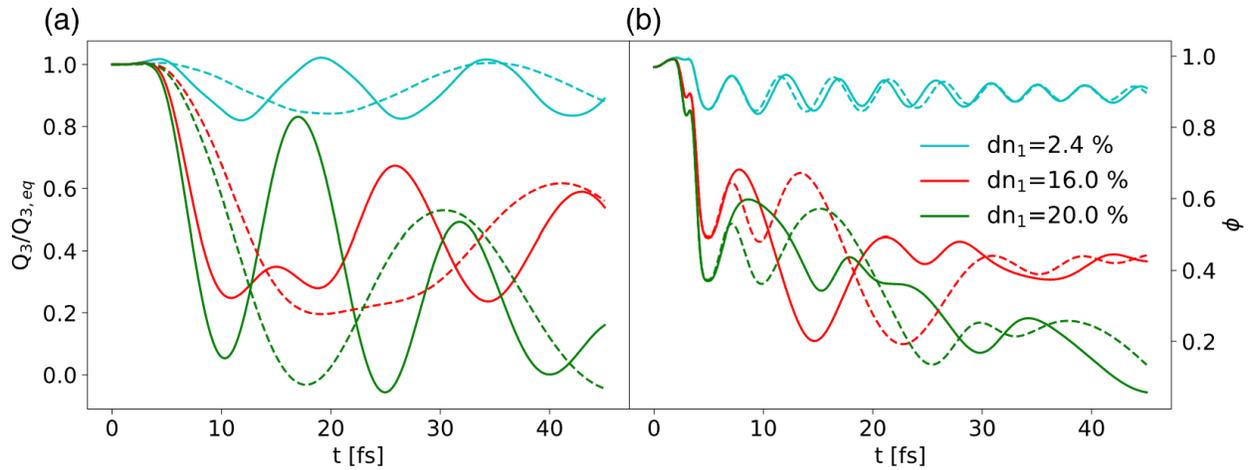

**Figure S10.** (a) Scaling of the short time evolution of the lattice mode $Q_3$. (b) Lattice disproportionation $\phi$ with a decreasing frequency of the lattice mode from $\omega_0$=250 meV (full lines) to $\omega_0$=125 meV (dashed lines).

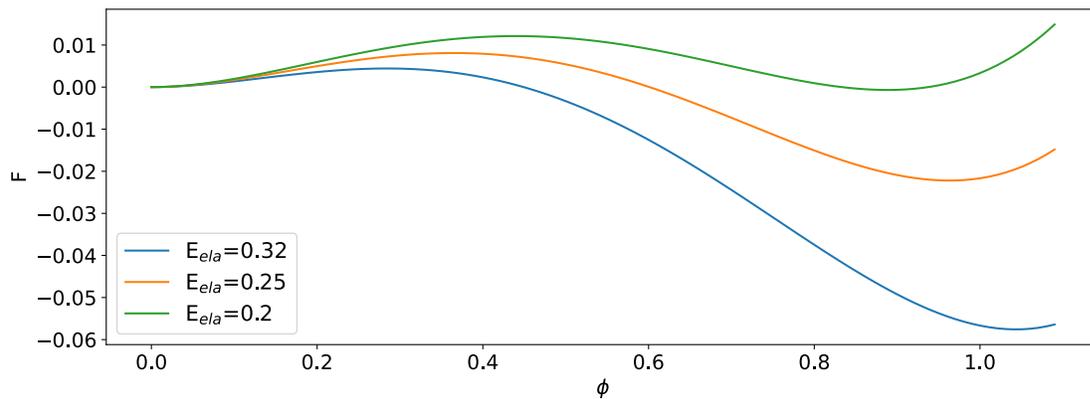

**Figure S11.** Free energy dependence on the electronic order parameter for various values of the lattice energy $E_{ela}=KQ_m^2/2$ [in eV].

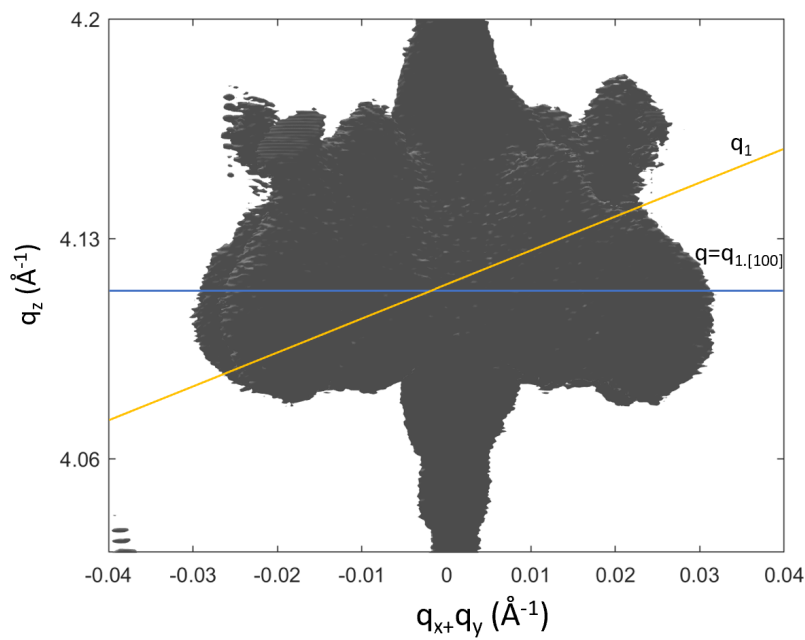

**Figure S12. Reciprocal space mapping of diffuse scattering intensity around the 008 Bragg peak.** The 3D rendering of the isosurface of the diffuse scattering and the 008 Bragg peak in the ground state. The yellow line shows the section of the Ewald sphere used to measure data. The scattering plane is within the [110] and [001] directions in Pbca orthorhombic coordinates. We define $q_x$ along **a**=[100], $q_y$ along **b**=[010], and $q_z$ along **c**=[001] in orthorhombic Pbca notation. The satellite peaks are rotated by 45 degrees around [001] with respect to the scattering plane. The data in Figs. 4d and S17 are shown as a function of $q$, which is the projection of $q_1$ onto $q_x$. This representation is chosen so the satellite peaks in the 3D reciprocal space (Fig. 4d a-c) and the time-resolved data (Fig. 4d) have the same momentum transfer.

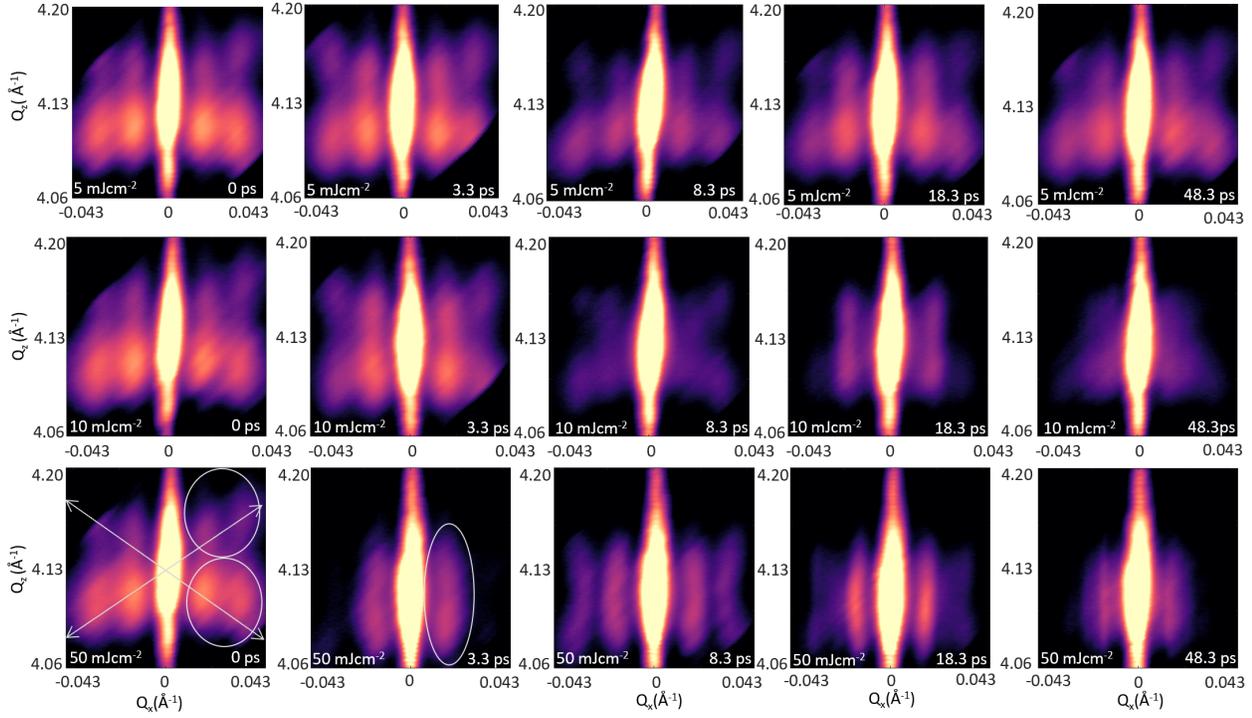

**Figure S13. Dynamics in the nano-texture inside the strained Ca$_2$RuO$_4$ thin film.** Time-resolved diffuse X-ray scattering of Ca$_2$RuO$_4$ shown on a q$_x$-q$_z$ slice through the 008 Bragg peak. We define q$_x$ along **a**=[100], q$_y$ along **b**=[010], and q$_z$ along **c**=[001] in orthorhombic Pbca notation. Satellite peaks along the q$_x$ direction signify a periodic nano-texture along [100] crystallographic direction (**a**-axis, oriented along the film surface). In the ground state, the periodicity of the nano-texture, $\Lambda_{[100]}$, can be estimated from the distance between the first-order satellite peak and the Bragg peak: $\Delta q_x \approx 0.015 \text{Å}^{-1}$, $\Lambda_{[100]} = 2 \cdot \pi / \Delta q_x \approx$ 40nm. The data is shown for various time delays and fluences. The fluences are indicated in the lower left corner, and time delays are indicated in the lower right corner of the false color images. In Fig c$_1$ (ground state), two circles indicate local maxima in both satellite peaks as a function of q$_z$. The satellites loosely follow a diagonal symmetry indicated by the white arrows. In Fig. c$_2$ and after, only one local maximum is visible, suggesting a change in the configuration of the interfaces.

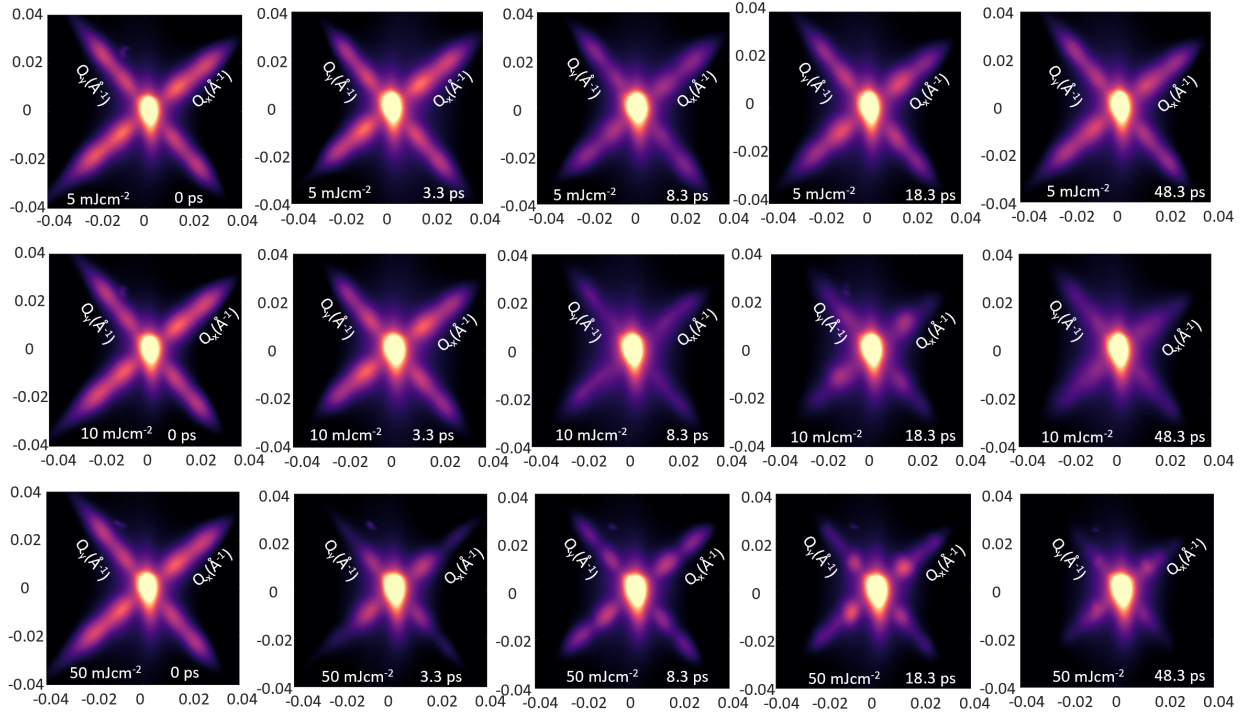

**Figure S14. Dynamics in the nano-texture inside the strained Ca$_2$RuO$_4$ thin film.** Time-resolved diffuse X-ray scattering of Ca$_2$RuO$_4$ around the 008 Bragg peak, projected onto the q$_x$-q$_y$ plane. We define q$_x$ along **a**=[100], q$_y$ along **b**=[010], and q$_z$ along **c**=[001] in orthorhombic Pbca notation. The data is shown for various time delays and fluences, indicated in the lower left and lower right corners of the false color images.

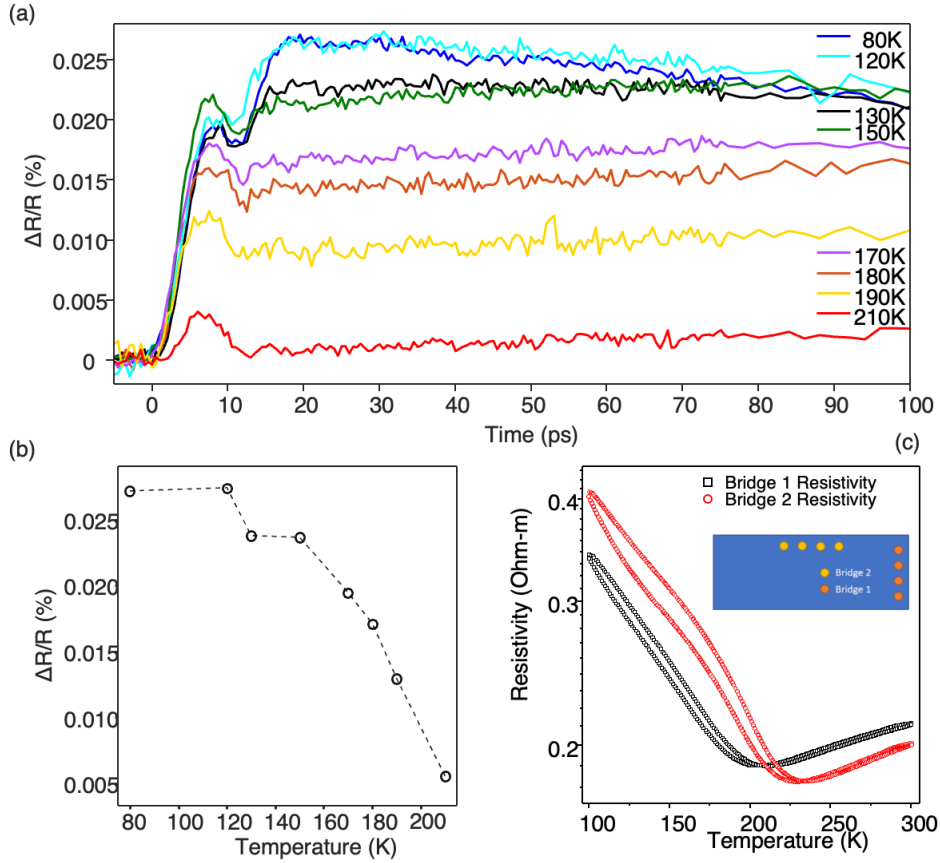

**Figure 15.** (a) THz reflectivity (data at 130 K is reproduced from Fig. 1c) measured as a function of temperature. The pump fluence is 10.5 mJ/cm$^2$. (b) The maximum change in reflectivity as a function of temperature. The change saturates at a temperature of 80-120 K. The saturation suggests that the THz reflectivity does not change appreciably below 80 K, which we interpret as the material near-fully transforming to an insulator at 80 K, and the material being close to fully transformed at 130 K. (c) Resistivity as a function of temperature measured along two orthogonal directions along the substrate edges on the strained $Ca_2RuO_4$ film.

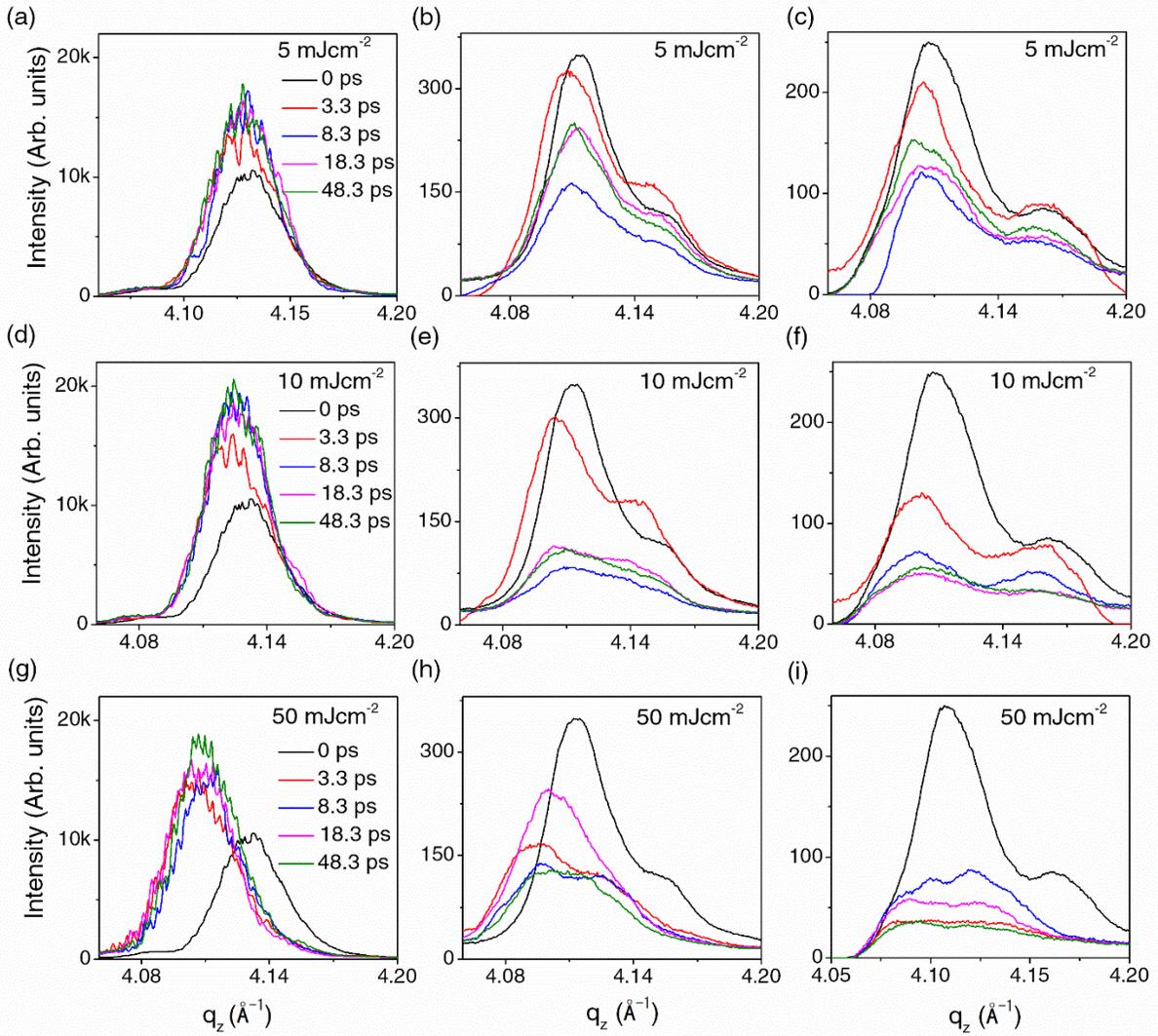

**Figure S16. The distribution of the satellite peak intensity along the $q_z$ direction as extracted from Figure S13.** The plots show data for (a-c) 5 mJcm$^{-2}$, (d-f) 10 mJcm$^{-2}$, and (g-i) 50 mJcm$^{-2}$. The three line scans were taken at $q_x=0$Å$^{-1}$ (left column, a,d,g), $q_x=0.015$Å$^{-1}$ (middle column, b,e,h), and $q_x=0.03$ Å$^{-1}$ (right column, c,f,i).

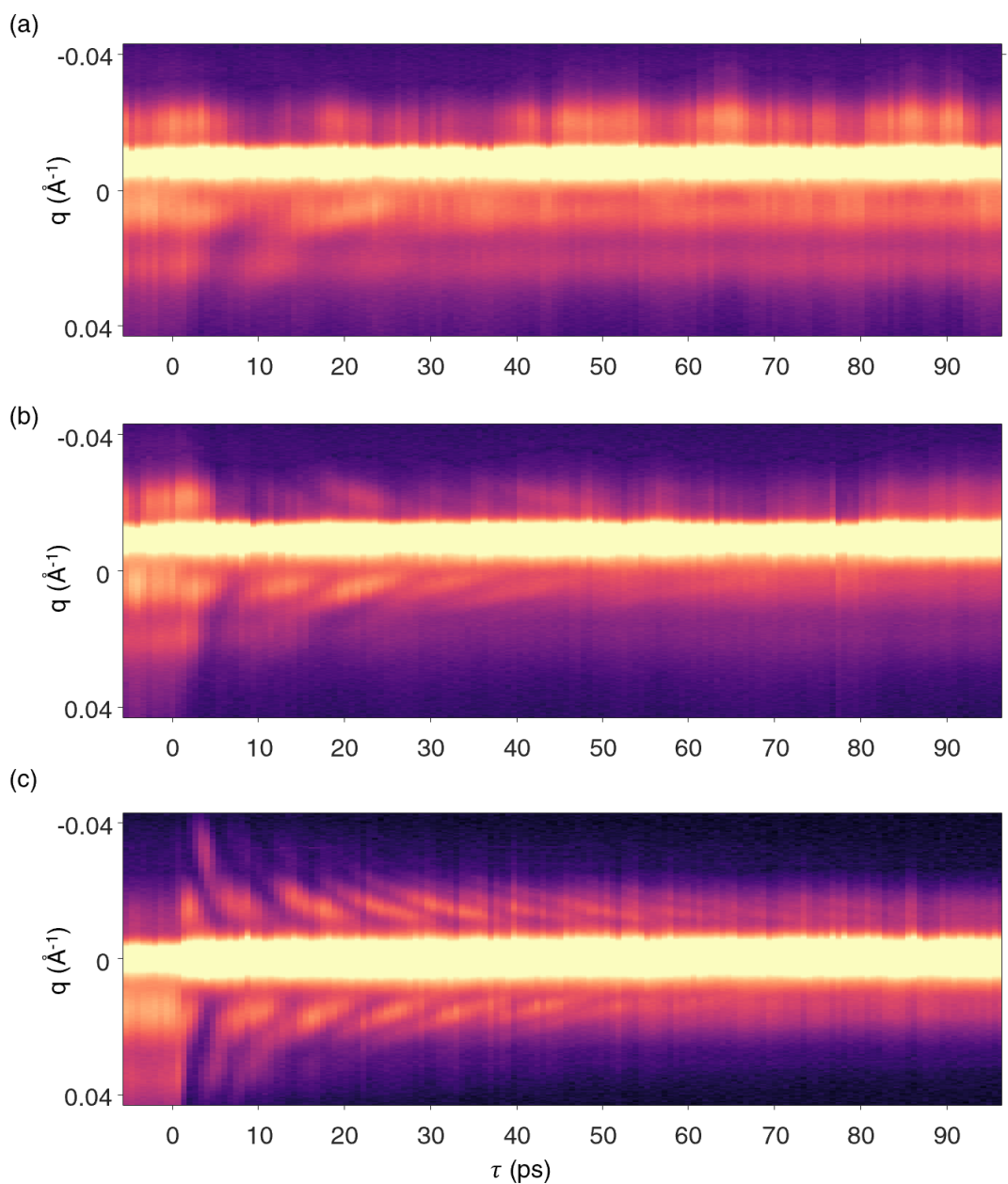

**Figure S17. Time-resolved diffuse X-ray scattering for different pump fluences.** Time-resolved reciprocal space map of 008 Bragg and satellite peaks of $Ca_2RuO_4$ at a pump fluence of (a) 5 mJcm$^{-2}$, (b) 10 mJcm$^{-2}$, and (c) 50 mJcm$^{-2}$, measured on a single slice of the Ewald sphere and projected onto $q_x$. Figure (c) is Figure 4d in the main text.

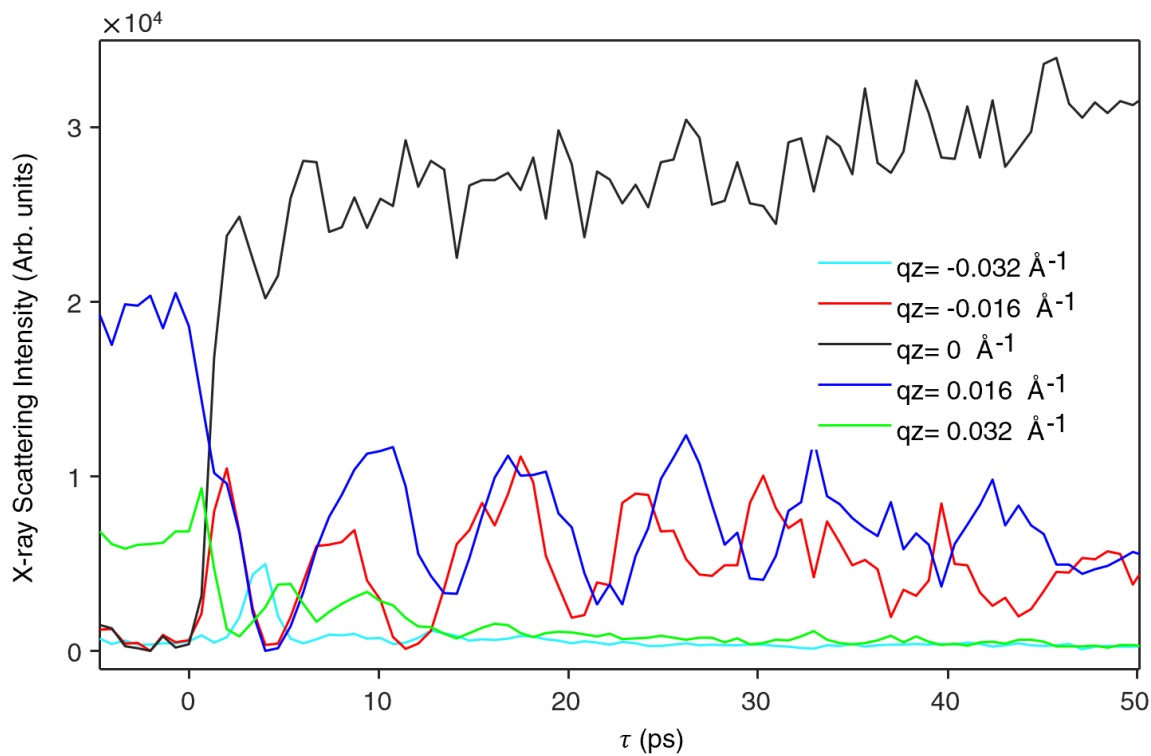

**Figure S18. The time-dependent diffuse scattering at different coordinates in reciprocal space.** Line profiles extracted from Fig. S17 at $q_x$=-0.032 Å$^{-1}$, -0.016 Å$^{-1}$, 0.00 Å$^{-1}$, 0.016 Å$^{-1}$, and 0.032 Å$^{-1}$ for 50 mJcm$^{-2}$. Note, the data are measured on a fixed Ewald sphere. Within the first 4 ps, the peak, including all satellites, shifts into the Ewald sphere: the dynamics within that time frame is a combination of peak shift and satellite dynamics. After 4 ps the shift is complete because the oscillations are smaller than the satellite peak width along $q_z$; the vibrations inside the crystal are mainly responsible for satellite peak dynamics. We parametrize the data as a projection on $q_x$, yet the Ewald sphere also has a large component along $q_z$. Both first-order and second-order satellite peaks at lower $q_z$ (also lower $q_x$), which corresponds to larger lattice constants, respond noticeably faster than the same satellite at higher $q_z$ (compare yellow and magenta line for first-order satellite peak and red and green line for the second-order satellite peak). Both second-order satellite peaks vanish within a few ps. For better visualization, all satellite peak intensities were multiplied by a factor of 100 as compared with data for $q_x$=0 Å$^{-1}$.

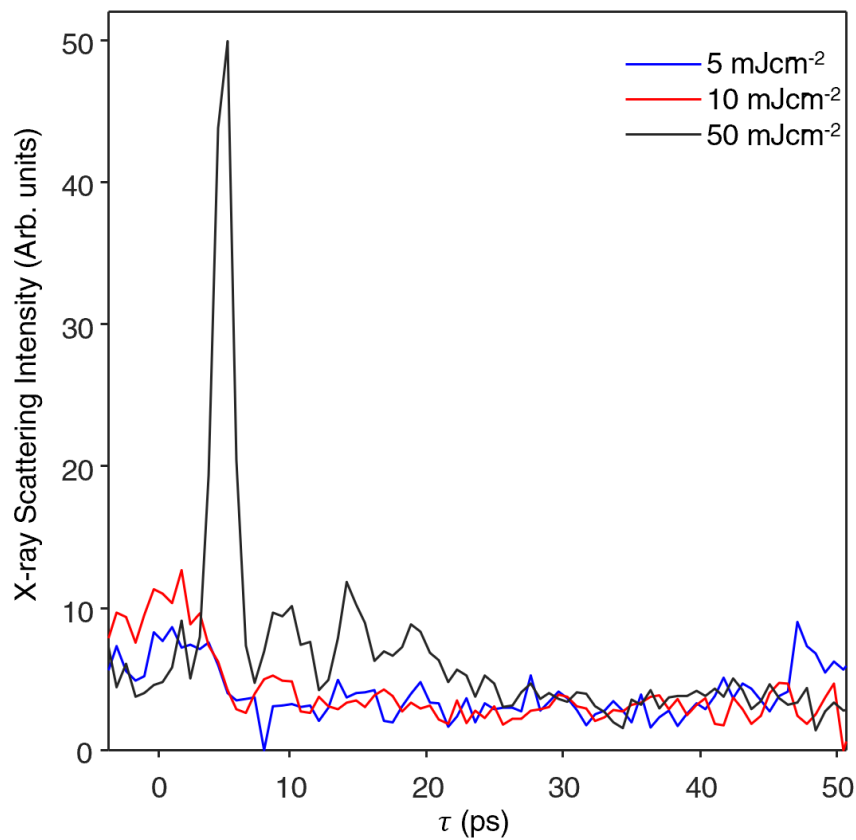

**Figure S19. The time-dependent diffuse scattering at a fixed point in reciprocal space.** Line profiles extracted from Figure S17 at $q_x$=-0.032 Å$^{-1}$. For 5 mJcm$^{-2}$ and 10 mJcm$^{-2}$, one observes a photoinduced reduction in diffuse scattering intensity, likely coincident with the lattice expansion. The diffuse scattering for a fluence of 50 mJcm$^{-2}$ shows a sharp peak at a time delay of ~5 ps after the change. This is not due to the peak shifting into the Ewald sphere, as that effect is visible in the first-order satellite peak and completes after 4 ps. The diffuse scattering could be due to the localized nature of the phase transformation, which generates diffuse scattering at a higher momentum transfer.

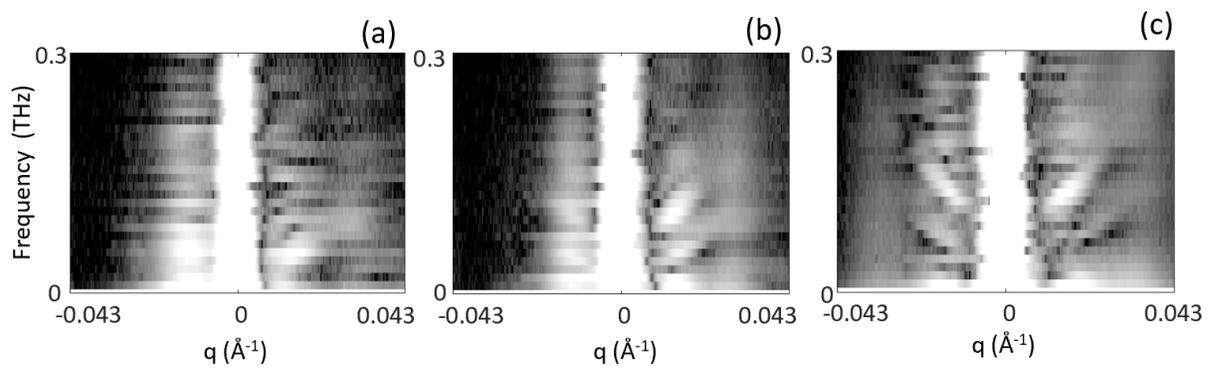

**Figure S20. Frequency dispersion relation for $Ca_2RuO_4$ thin film**. The frequency spectrum is extracted from data shown in Figure S17 via the Fast Fourier Transform of time-resolved data at different pump fluences, where (a) 5 mJcm$^{-2}$, (b) 10 mJcm$^{-2}$, and (c) 50 mJcm$^{-2}$.

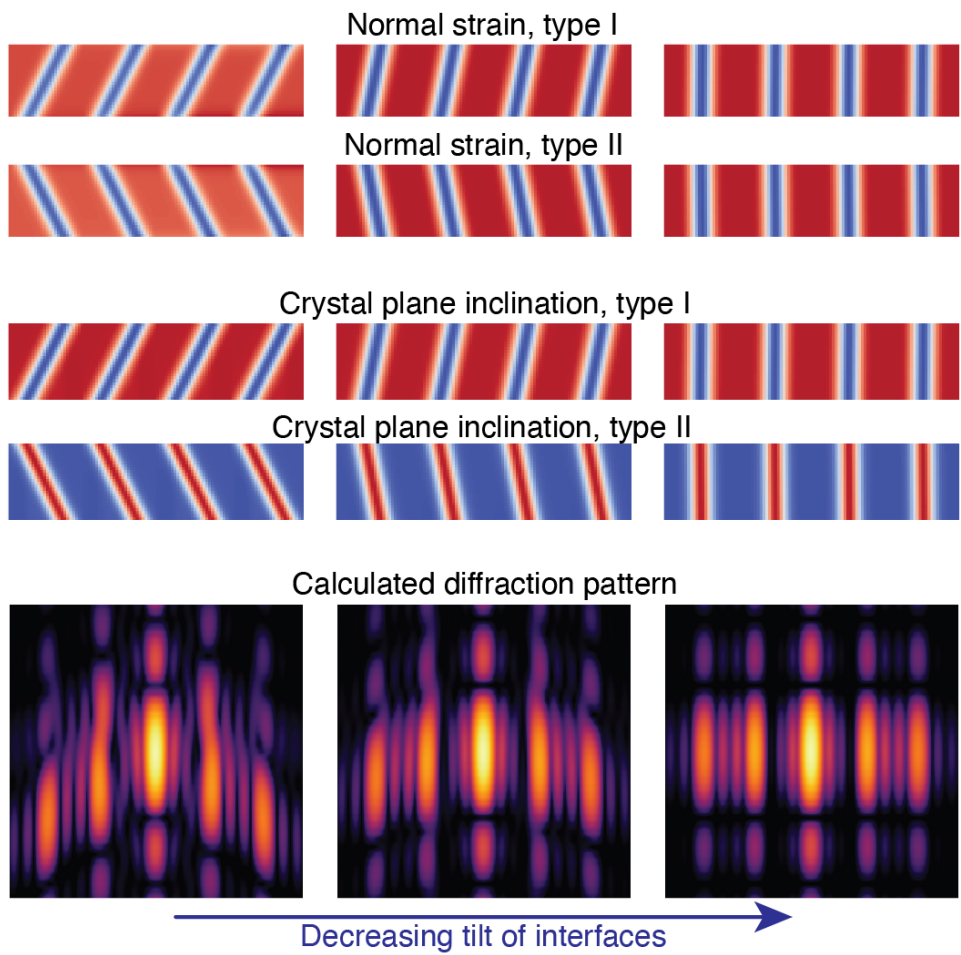

**Figure S21. X-ray modeling of the nanotexture.** The simulated displacement field from two different morphologies with nanotexture tilted to the right (type I) and to the left (type 2). Both morphologies are present in the film because of the square symmetry of the substrate surface (the same morphology is present with a rotation of 90 degrees). Both derivatives of the displacement field are shown, the normal strain and the crystal inclination (red is positive and blue is negative). The displacement field was calculated as an integral of its two derivatives. The diffraction patterns are calculated as a sum of the two intensities calculated from the displacement field of type I and type II. The diffraction is asymmetric around the vertical axis for tilted interfaces (similar to the low-temperature ground state shown in Fig. 4a) and symmetric for untitled interfaces (similar to the photo-excited state shown in Fig. 4c). The colormap in the diffraction pattern is identical to Figure 4.

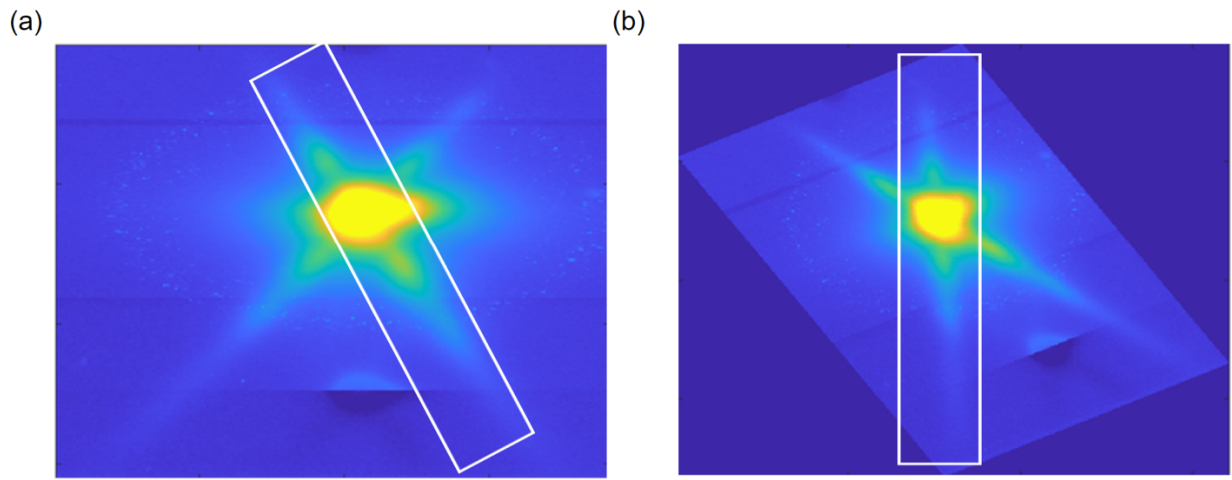

**Figure S22.** (a) A typical diffraction profile collected at the XFEL. A series of such images were transformed into a 3D reciprocal space map (see Fig. S12). Fig. 4 a-c in the main paper shows sections of the transformed reciprocal space map, taken along the cartesian coordinates $q_x$, $q_y$, $q_z$. (b) For faster time dynamics shown in Fig. 4d, we rotated the diffraction data as shown in (b) and averaged the signal in the white box in the horizontal direction.